\documentstyle[prl,aps,preprint,amsfonts]{revtex}
\newcommand{\tr}{{\rm tr}}
\newcommand{\Tr}{{\rm Tr}}
\newcommand{\Dirac}{{\rm D}}

\newcommand{\thru}[1]{\mathrel{\mathop{#1\!\!\!/}}}
\newcommand{\eq}[1]{eq.~(\ref{eq:#1})}

\newcommand{\D}{D}
\newcommand{\Da}{\hat{D}}

\newcommand{\Ar}{{\cal A}}

\newcommand{\sm}{\mbox{\textsf{\textit{m}}}}

\newcommand{\sv}{\mbox{\textsf{\textit{v}}}}

\newcommand{\sD}{\mbox{\textsf{\textit{D}}}}
\newcommand{\sDa}{\hat{\sD}}
\newcommand{\sF}{\mbox{\textsf{\textit{F}}}}
\newcommand{\sJ}{\mbox{\textsf{\textit{J}}}}
\newcommand{\sL}{\mbox{\textsf{\textit{L}}}}
\newcommand{\sM}{\mbox{\textsf{\textit{M}}}}
\newcommand{\sP}{\mbox{\textsf{\textit{P}}}}
\newcommand{\sR}{\mbox{\textsf{\textit{R}}}}
\newcommand{\sU}{\mbox{\textsf{\textit{U}}}}
\newcommand{\sX}{\mbox{\textsf{\textit{X}}}}

\newcommand{\mlr}{m_{LR}}
\newcommand{\mrl}{m_{RL}}

\newcommand{\m}[1]{{\underline{#1}}}

\begin{document}

\draft
\tighten
\def\footnoterule{\kern-3pt \hrule width\hsize \kern3pt}

\title{Derivative expansion for the effective action of
chiral gauge fermions. The abnormal parity component.}

\author{L.L. Salcedo}

\address{
{~} \\
Departamento de F\'{\i}sica Moderna \\
Universidad de Granada \\
E-18071 Granada, Spain
}

\date{\today}
\maketitle

\thispagestyle{empty}

\begin{abstract}
Explicit exact formulas are presented, for the leading order term in a
strict chiral covariant derivative expansion, for the abnormal parity
component of the effective action of two- and four-dimensional Dirac
fermions in presence of scalar, pseudo-scalar, vector and axial vector
background fields. The formulas hold for completely general internal
symmetry groups and general configurations. In particular the scalar
and pseudo-scalar fields need not be on the chiral circle.
\end{abstract}

\vspace*{1cm}
PACS numbers:\ \ 11.30.Rd 11.15.Tk 11.10.Kk

\vspace*{0.2cm}
Keywords:\ \ chiral fermions, fermion determinant,
derivative expansion, effective action, gauge field theory, anomaly

\vspace*{\fill}

\section{Introduction}
\label{sec:1}

This paper is the second part of the work initiated in
\cite{Salcedo:2000I} on the explicit computation of the effective
action of chiral gauge fermions, including scalar, pseudo-scalar,
vector and axial vector external fields, within a strict covariant
derivative expansion. Ref. \cite{Salcedo:2000I} dealt with the real
part of the effective action and here the imaginary part is worked out
at leading order for two- and four-dimensional fermions. The main
feature of both works is that a strict covariant derivative expansion
is carry out, rather than a perturbative, commutator or heat-kernel
expansion, and that explicit formulas are given which hold without
putting any restrictions on the external field configurations nor
making assumptions on the internal symmetry group. In fact this
generality helps to concentrate on the computational issues and
results in a easier calculation.

The imaginary part of the effective action of chiral gauge fermions
(the phase of the fermionic determinant) displays some well-known
peculiarities as compared to the real part. It presents a $2\pi i$
multivaluation, anomalies in the chiral symmetry and contains
topological pieces. In comparison the real part only displays a scale
anomaly, which however is absent in the imaginary part. These
peculiarities make this piece more interesting from the theoretical
point of view and has been the source of deeply original insights
\cite{Adler:1969av,Bell:1969re,Bardeen:1969md,Adler:1969er,Wess:1971cm,Fujikawa:1979ay,Goldstone:1981kk,Witten:1983tw,Bardeen:1984pm}.
Consequently it has been extensively studied in the literature (for
reviews see e.g. \cite{Ball:1989xg,Alvarez-Gaume:1985dr}.)

The presence of the chiral anomaly introduces some mathematical
subtleties in the definition of the effective action at the
non-perturbative level \cite{Ball:1989xg} since the chirally covariant
renormalized current (the variation of the effective action) fails to
be consistent \cite{Leutwyler:1985em}. These complications are also
present in the computation of the effective action in the framework of
an asymptotic expansion, such as the covariant derivative expansion to
be considered here. A direct computation must necessarily break chiral
invariance and becomes prohibitive if one insist on a strict
derivative expansion except for particular internal symmetry
groups. The reason is that in a strict covariant derivative expansion
both the scalar and the pseudo-scalar fields must be treated
non-perturbatively, and as a rule it not possible to treat two or more
operators non-perturbatively unless they commute. For instance in
\cite{Salcedo:1996qy} such a calculation is done for two-dimensional
fermions with SU(2) internal symmetry group. In that case the
particular algebraic properties of su(2) allowed to carry out the
computation, but the same method cannot be extended to general groups.

An alternative method is to make a chiral rotation to fix the chiral
gauge so that there is no pseudo-scalar field. Then a direct
calculation becomes possible, using for instance a $\zeta$-function
approach combined with a symbols method \cite{Salcedo:1996qy}. Because
the chiral gauge has been fixed (or rather, reduced to a manageable
vector gauge invariance), such as result is, in some sense, manifestly
chiral gauge invariant (the anomaly comes through the
Wess-Zumino-Witten term generated by the chiral rotation). However,
this procedure is not completely satisfactory for various reasons. The
result would be given in terms of the rotated variables rather than in
terms of the original external fields. In addition, it does not fully
exploit the symmetries of the problem; as will be shown, within the
derivative expansion the effective action depends analytically on the
external fields, in a sense to be made more precise below, and this
property is not explicit in terms of the rotated
variables. Analyticity is a property of the effective action
functional which is not shared by most functionals that are chiral
invariant (modulo anomalies). Another important shortcoming of that
method, as compared to the method to be presented here, is that the
functional depends on three objects, the (rotated) scalar field $S$,
the axial field $A$ and the vector gauge covariant derivative $D_V$,
whereas in our approach there are just two objects, an effective
scalar field $\sm$ and an effective vector gauge covariant derivative
$\sD$ which behave almost as those of a vector-like theory (i.e., a
theory without pseudo-scalar nor axial vector fields). This results in
a great reduction of the amount of algebra required, due to the
smaller number of algebraic combinations and the fact that analyticity
is preserved throughout.

The method proposed in this work is based on using a suitable notation
which allows to map certain chiral invariant objects with an
analytical form (and in particular the chiral covariant effective
current) to the corresponding object in an effective vector-like
theory. This allows to carry out computations in the effective
vector-like theory and then map back the result into the chiral
setting. This is a kind of analytical extension from the vector-like
case to the full chiral case and the chiral result so obtained
corresponds to the LR version of the effective action. This procedure
is very convenient from the computational point of view since the
vector-like case is very well understood and many methods exists to
dealt with it. In particular the issue of renormalization is almost
trivial since the requirement of (effective) vector gauge invariance
completely fixes the form of the effective action.

For the normal parity component the mapping between chiral and vector
theories is literal and applies to the effective action itself. This
is exploited in \cite{Salcedo:2000I}. In the abnormal parity sector
the mapping from vector to chiral holds whenever the trace cyclic
property is not involved, e.g. for the covariant effective current,
but not for the effective action itself since this would not allow for
the existence of the chiral anomaly. In our formalism this is
reflected in the fact that $\sm$ is odd under cyclic
transformations. Our strategy will then be along the lines of
Schwinger's method \cite{Schwinger:1951nm}, i.e., we first to compute
the covariant effective current and subsequently use this current to
recover the effective action. This second step is done by writing down
an explicit analytical functional of the type of the
Wess-Zumino-Witten term which saturates the chiral symmetry breaking
terms of the effective action, and adjusting the remainder, which
necessarily will be chiral invariant, so that the correct current is
reproduced. The calculation of the current is done from scratch by
using essentially the method of symbols, but in the improved version
due to Pletnev and Banin \cite{Pletnev:1999yu} which reduces the
amount of algebra while preserving explicit gauge invariance
throughout.
 
In Section \ref{sec:2} we recall the notation introduced in
\cite{Salcedo:2000I} and extend it to cover the abnormal parity case.
A set of notational conventions are introduced so that the chiral case
can, to a large extent, be treated as a vector theory. A further
convention is introduced which allows to carry out explicit loop
momentum integrations without assuming commutativity of the operators
involved. This convention is illustrated in the same section with the
Wess-Zumino-Witten action which is brought into an explicit Lagrangian
form preserving manifest global vector gauge invariance, for a general
gauge group. In Section \ref{sec:4} the chiral covariant effective
current is explicitly computed at leading order in the derivative
expansion using the method of symbols for the two- and
four-dimensional cases. In Section \ref{sec:3} we introduce an
extended version of the gauged Wess-Zumino-Witten action which holds
off the chiral circle and depends analytically on the external
fields. Next we consider the general form of the possible chiral
invariant remainder (which saturates the full abnormal parity
effective action at leading order). This remainder is then explicitly
determined from the current.  In Section \ref{sec:5} several comments
and extensions are given. In subsection \ref{subsec:V.a} we show that
on the chiral circle our extended gauge Wess-Zumino-Witten term
reduces to the usual one and the chiral invariant remainder
vanishes. This result is extended to the case of an Abelian chiral
radius and in particular to the full Abelian case. In subsection
\ref{subsec:V.e} the effective density (the variation of the effective
action with respect the scalar and pseudo-scalar fields) is explicitly
computed and the anomalous continuity equation verified.  Both for the
current and for the density an unexpected extra symmetry is found
which does not follow from Lorentz and chiral symmetries but seems to
depend on the concrete properties of the effective action
functional. In subsection \ref{subsec:V.f} the VA version of the
effective action is considered and the corresponding formulas are
given for the particular case of vanishing pseudo-scalar field.  In
subsection \ref{subsec:V.d} we show that the imaginary part of the
effective action vanishes when one of the matter chiral fields is a
spacetime constant and there are no chiral gauge fields. Next we show
how this observation, plus the assumption of analyticity, is
sufficient to completely determine the effective action in two
dimensions and puts restrictions in higher dimensions.  In subsections
\ref{subsec:V.c} and \ref{subsec:V.b} we consider further properties
of the extended gauged Wess-Zumino-Witten term and of the chiral
invariant remainder. Finally in subsection \ref{subsec:V.h} we verify
a descent relation which relates the effective action in $d$
dimensions with the vector current in $d+2$ dimensions and the
Chern-Simons term in $d+1$ dimensions. In Appendix \ref{app:A} we
collect the formulas corresponding to the chiral anomaly and the
various versions of the Wess-Zumino-Witten action and Appendix
\ref{app:B} contains the explicit formulas for the effective action
and the effective current in two and four dimensions.

\section{Notation and conventions}
\label{sec:2}

We will follow the notation and conventions summarized in Section II
of \cite{Salcedo:2000I}. The extensions needed to adapt these
conventions to the pseudo-parity odd case are presented
below. (Ref. \cite{Salcedo:2000I} deals with the pseudo-parity even
component of the effective action, $W^+$.)  Because some of these
conventions are not standard, the reader is invited to consult the
Section II of \cite{Salcedo:2000I} for further details.

The spacetime is Euclidean and flat and its dimension $d$ is even. The
class of Dirac operators to be considered is
\begin{equation}
\Dirac= \thru{\D}_R P_R +\thru{\D}_LP_L + \mlr P_R+\mrl P_L \,
\end{equation}
where $P_{R,L}=\frac{1}{2}(1\pm\gamma_5)$ are the projectors
on the subspaces $\gamma_5=\pm 1$. Our conventions are
\begin{equation}
\gamma_\mu=\gamma_\mu^\dagger\,,\quad \{\gamma_\mu,\gamma_\nu\}=
2\delta_{\mu\nu}\,,\quad
\gamma_5=\gamma_5^\dagger=\gamma_5^{-1}=
\eta_d\gamma_0\cdots\gamma_{d-1}\,,\quad
\tr_{\rm Dirac}(1)=2^{d/2}\,,
\label{eq:v6}
\end{equation}
where $\eta_d= \pm i^{d/2}$ (a concrete choice will not be needed
except in Section \ref{subsec:V.h}).  $\D^{R,L}_\mu=
\partial_\mu+v^{R,L}_\mu$ are the chiral covariant derivatives. The
external bosonic fields $v^{R,L}_\mu(x)$ and $\mlr (x)$, $\mrl (x)$
are matrices in some generic internal space (referred to as flavor),
the identity in Dirac space and multiplicative operators in $x$
space. In order to avoid infrared divergences we will assume that the
matrices $\mlr$ and $\mrl$ are nowhere singular. No algebraic
assumptions will be made on the internal space matrices in the
derivation of our results. Of course, at the end, they can be applied
to particular interesting cases, such as Abelian groups, or the case
of scalar fields on the so-called chiral circle, $\mlr(x)\mrl(x)=M^2$,
$M^2$ being a constant c-number, for which many results exists.

In what follows, the symbol $\langle\ \rangle$ will we used as a
short-hand to denote
\begin{equation}
\langle X \rangle_{d,B} = 
\frac{\eta_d (d/2)!}{(2\pi)^{d/2}d!}\int_B\tr(X)\,.
\label{eq:n2}
\end{equation}
In this formula $d$ is the space-time dimension, $\eta_d$ is the
normalization in $\gamma_5$, $\tr$ refers to flavor only, $B$ is some
$n$-dimensional integration region, and $X$ is some differential
$n$-form which is a matrix in flavor space. In general $B$ will be the
spacetime and $X$4 a $d$-form, and the subscripts $B$ and $d$ will be
suppressed.

\subsection{Specific conventions}

The effective action is a functional of the external fields, defined
as $W[v,m]=-\Tr\log(\Dirac)$, where some regularization plus
renormalization is understood.  The pseudo-parity transformation is
defined as the operation of exchanging the chiral labels $R$ and $L$
everywhere. The effective action then decomposes naturally into a
pseudo-parity even (or normal parity) component, $W^+[v,m]$, and a
pseudo-parity odd (or abnormal parity) one, $W^-[v,m]$. The latter is
also characterized by being purely imaginary (in Euclidean space),
containing the Levi-Civita pseudo-tensor, having topological pieces,
displaying multivaluation by integer multiples of $2\pi i$, and
presenting an anomaly under chiral transformations.

The effective action can be expanded into terms with a well-defined
number of covariant derivatives (or equivalently, of Lorentz
indices). For each such term $T$, one can consider its pseudo-parity
conjugate $T^*$, i.e., the same expression as $T$ after the exchange
of all labels $L$ with $R$. Then we will adopt the following
convention (see Section II of \cite{Salcedo:2000I} for further details):
\begin{description}
\item[Convention 1.] In $W^+$, the terms $T$ and $T^*$ will be
identified, so that under this convention $T$ actually stands for
$\frac{1}{2}(T+T^*)$. In $W^-$ every term $T$ is identified with
$-T^*$ and thus $T$ stands for $\frac{1}{2}(T-T^*)$.
\end{description}

Consider now a typical chiral invariant expression such as
$\tr(F_{\mu\nu}^R \Da_\mu \mrl \Da_\nu \mlr)$.  (As usual,
$$
\Da_\mu\mrl=\D^R_\mu\mrl-\mrl\D^L_\mu, \quad
F_{\mu\nu}^R=[\D_\mu^R,\D_\nu^R]\,,
$$ etc.)  It can be observed that each factor falls into one of the
following classes, according to its chiral labels, namely $RR$, $LL$,
$RL$ and $LR$.  For instance, $\Da_\nu\mlr$ lies in the class $LR$. By
inserting such a factor in an expression, the chiral label is flipped
from $R$ to $L$ as one moves from right to left in the formula (or
equivalently on the fermion loop). On the other hand $F^R_{\mu\nu}$
belongs to the class $RR$ and it does not flip the chiral label.
Further, it is observed that in such a chiral invariant expression any
two adjacent chiral labels belonging to two different factors are
equal (e.g. the label $L$ in $\Da_\mu \mrl \Da_\nu \mlr$). This must
be so in order to preserve covariance under chiral
transformations. Moreover, if the expression is inside the trace, the
first and last chiral labels must also coincide for the same reason,
due to the cyclic property of the trace. Thus in chiral covariant
expressions the following convention can be used (see Section II of
\cite{Salcedo:2000I} for further details):
\begin{description}
\item[Convention 2.] In expressions where the chiral labels are
combined preserving chirality, these labels are redundant and will be
suppressed, so a term such as $X_{RR}Y_{RL}Z_{LR}$ will be written as
$(\textsf{\textit{XYZ}})_{RR}$.\footnote{This example assumes, of
course, that we know beforehand that $Y$ and $Z$ flip the chirality
label and $X$ does not. This is the case in practice since, $W^\pm$ is
constructed with $\D^{R,L}_\mu$, $\mlr$ and $\mrl$.} Inside a trace it
is sufficient to write $\tr(\textsf{\textit{XYZ}})$ plus the
convention that the first (and last) implicit label is $R$. (This last
convention is needed to fix the sign in the pseudo-parity odd case.)
\end{description}
For instance
\begin{eqnarray}
\tr(\sF_{\mu\nu}\,\sDa_\mu\sm\,\sDa_\nu\sm) &=&
\tr(F_{\mu\nu}^R \Da_\mu \mrl \Da_\nu \mlr)
\nonumber \\&=&
\pm\tr(F_{\mu\nu}^L \Da_\mu \mlr \Da_\nu \mrl)
\nonumber \\
 &=&
\frac{1}{2}\tr(F_{\mu\nu}^R \Da_\mu \mrl \Da_\nu \mlr) \pm
\frac{1}{2}\tr(F_{\mu\nu}^L \Da_\mu \mlr \Da_\nu \mrl)\,.
\end{eqnarray}
The $\pm$ refers to $W^\pm$ respectively.

In the pseudo-parity even sector, the cyclic property of the trace
works as usual within the index-free notation introduced by Convention
2 \cite{Salcedo:2000I}. However, the cyclic property is modified for
$W^-$. Let $X$ be of type $LR$ or $RL$ then
\begin{equation}
\tr(\sX\sm)= \tr(X_{RL}\mlr)= \tr(\mlr X_{RL})= \pm\tr(\mrl X_{LR})=
\pm\tr(\sm\sX)\,,\quad\text{in $W^\pm$}\,.
\label{eq:10}
\end{equation}
This is equivalent to saying that, in $W^-$, the object $\sm$ changes
sign under the cyclic property. The same is true for any object that
flips the chiral label, i.e. of the type $RL$ or $LR$. Consider now
the following identities in $W^-$ (where $f$ and $g$ are ordinary
functions)
\begin{equation}
\tr(f(\sm)g(\sm))= \tr(g(\sm)f(-\sm))=\tr(g(-\sm)f(\sm))\,.
\end{equation}
The first equality follows from moving $f(\sm)$ to the right, the
second one from moving $g(\sm)$ to the left using the (modified)
cyclic property. This equality implies that only the even component of
the function $f(x)g(x)$ (under $x\to -x$) contributes. This is just an
illustration of the obvious consistency condition stating that the
number of chirality flipping factors must always be even
(e.g. $f(\sm)g(\sm)$ must contain even powers of $\sm$ only) because
in any expression inside the trace the first and the last chiral
labels must coincide due to chiral invariance. This observation
applies to $W^+$ as well.

In this notation, the chiral rotations
$\mlr\to\Omega_L^{-1}\mlr\Omega_R$, etc, become
\begin{equation}
\sm \to \Omega^{-1}\sm\Omega \,, \quad 
\sv_\mu \to \Omega^{-1}\sv_\mu\Omega+\Omega^{-1}\partial_\mu\Omega\,,
\end{equation}
whereas for infinitesimal rotations, $\Omega_{R,L}=
\exp(\alpha_{R,L})$ with $\alpha_{R,L}$ infinitesimal
\begin{equation}
\delta\sm = [\sm,\alpha] \,, \quad \delta\sv_\mu = \sDa_\mu\alpha\,.
\label{eq:7b}
\end{equation}

A further convention is introduced in \cite{Salcedo:2000I}, namely
\begin{description}
\item[Convention 3.] In an expression $f(A_1,B_2,\dots)XY\cdots$ the
ordering labels $1,2,\dots$ will denote the actual position of the
operators $A,B,\dots$ relative the fixed elements $X,Y,\dots$ so that
$A$ is to be placed before $X$, $B$ between $X$ and $Y$, etc. That
is, for a separable function $f(a,b,\dots)=\alpha(a)\beta(b)\cdots$,
the expression stands for $\alpha(A)X\beta(B)Y\cdots$
\end{description}
Note that this convention is independent of Conventions 1 and
2. Combining the several conventions and the cyclic property one has,
for instance
\begin{equation}
\tr(f(\sm_1,\sm_2)\sF_{\mu\nu}\sF_{\alpha\beta})=
\tr(f(-\sm_3,\sm_2)\sF_{\mu\nu}\sF_{\alpha\beta})=
\tr(f(-\sm_2,\sm_1)\sF_{\alpha\beta}\sF_{\mu\nu})
\label{eq:v2}
\end{equation}
in $W^-$. In the first equality, $\sm_1$ ($\sm$ in position 1) is
moved to position 3 (i.e. becomes the rightmost factor) using the
cyclic property, becoming $-\sm_3$. Then, in the second equality
$\sF_{\mu\nu}$ is moved to the rightmost position, and the position
labels of $\sm$ are modified accordingly.

Before proceeding, let us comment on the meaning of an expression,
such as $f(A_1,B_2,C_3)XY$, with operators acting in
different positions. It should be clear such an operator is a
well-defined one. The simplest way to reduce it to a more usual form
is by expressing the function $f$ as a linear combination of
separable functions,
\begin{equation}
f(z_1,z_2,z_3)= \sum_i\alpha_i(z_1)\beta_i(z_2)\gamma_i(z_3)\,,
\end{equation}
then
\begin{equation}
f(A_1,B_2,C_3)XY= \sum_i\alpha_i(A)\,X\,\beta_i(B)\,Y\,\gamma_i(C)\,,
\end{equation}
and the right-hand side is perfectly well-defined. In fact such a
representation in terms of separable functions is the usual means by
which the Convention 3 enters in the calculations (typically the sum
over $i$ corresponds an integration over the momentum of the loop).
An alternative method to fully characterize the operator
$f(A_1,B_2,C_3)XY$ is by means of its matrix elements. In this
context, the natural procedure is to use as basis the ones formed by
the eigenvectors of the operators $A$, $B$ and $C$. Let us denote
these basis by $|n,A\rangle$, $|m,B\rangle$ and $|r,C\rangle$, with
associated eigenvalues $a_n$, $b_m$ and $c_r$, and let $\langle n,A|$,
$\langle m,B|$ and $\langle r,C|$ the corresponding dual basis, then
\begin{equation}
\langle n,A| f(A_1,B_2,C_3)XY|r,C\rangle= 
\sum_m f(a_n,b_m,c_r) X_{nm}Y_{mr}, 
\end{equation}
where $X_{nm}=\langle n,A| X|m,B\rangle$ and $Y_{mr}=\langle
m,B|Y|r,C\rangle$. (This is easily established using the previous
representation in terms of separable functions.) This kind of
representation is in fact the one usually employed in the literature
(see e.g. \cite{Ball:1989xg}). The point to be emphasized is that the
operator depends solely on the function $f$ itself and not on any
particular representation.

Special care requires the use of the Convention 3 in combination with
Conventions 1 and 2 in practical applications. This is because the
meaning of the symbols under the Convention 2, depends on its position
in the formula. For instance, in an expression such as
$\tr(f(\sm_1,\sm_2)\sF_{\mu\nu}\sF_{\alpha\beta})$, where $f$ is a
complicated function it may not be clear how to expand the formula,
i.e., how to put back the chiral labels. Fortunately, there is a
simple general procedure to do so, namely, to decompose $f$ into its
even and odd components under $\sm_{1,2}\to\pm\sm_{1,2}$,
\begin{equation}
f(\sm_1,\sm_2)=
A(\sm_1^2,\sm_2^2)+\sm_1B(\sm_1^2,\sm_2^2)+\sm_2C(\sm_1^2,\sm_2^2)+
\sm_1\sm_2D(\sm_1^2,\sm_2^2)\,.
\label{eq:24a}
\end{equation}
As noted above, consistency requires $f$ to be even under
$\sm\to-\sm$, thus $B=C=0$. This produces
\begin{eqnarray}
\tr[f(\sm_1,\sm_2)\sF_{\mu\nu}\sF_{\alpha\beta}]
&=&
\tr[A(\sm_1^2,\sm_2^2)(\sF_{\mu\nu})(\sF_{\alpha\beta})]
+\tr[D(\sm_1^2,\sm_2^2)(\sm\sF_{\mu\nu})(\sm\sF_{\alpha\beta})] \,.
\label{eq:n24}
\end{eqnarray}
Now, from our conventions it unambiguously follows that the chiral
labeling is
\begin{eqnarray}
\tr[A(m^2_{R1},m^2_{R2})
(F_{\mu\nu}^R)(F_{\alpha\beta}^R)]
+\tr[D(m^2_{R1},m^2_{L2})
(\mrl F_{\mu\nu}^L)(\mlr F_{\alpha\beta}^R)] \,.
\label{eq:n24a}
\end{eqnarray}
In this formula Convention 1 still applies, $m^2_R=\mrl\mlr$ and
$m^2_L=\mlr\mrl$.\footnote{Alternatively, the second term could have
been written as
$\tr[D(\sm_1^2,\sm_2^2)(\sm\sF_{\mu\nu}\sm)(\sF_{\alpha\beta})]$,
yielding $\tr[D(m^2_{R1},m^2_{R2}) (\mrl F_{\mu\nu}^L\mlr)(
F_{\alpha\beta}^R)]$. This is equivalent to the previous result.}

Several illustrations of the Convention 3 (besides its use in $W^+$)
have been presented in \cite{Salcedo:2000I} and elsewhere
\cite{Garcia-Recio:2000gt,Salcedo:1999sv}. Here we present another
application which will be needed below. First let us introduce a
standard differential geometry notation: the quantities $dx_\mu$ are
anticommuting, $d^dx=dx_0dx_1\cdots dx_{d-1}$, $d$ is the differential
operator $dx_\mu\partial_\mu$, $\sv$ stands for $\sv_\mu dx_\mu$,
$\sD=\sD_\mu dx_\mu$, $\sF=\sD^2=d\sv+\sv^2$, etc. Consider now the
following $n$-form
\begin{equation}
X = f(A_1,\dots,A_n)(dA)^n\,,
\label{eq:v1}
\end{equation}
where $A$ is some matrix-valued function defined on some manifold, and
$f(z_1,\dots,z_n)$ is an ordinary function. We want to compute $dX$.
To this end, recall the rule \cite{Salcedo:2000I}
\begin{equation}
\delta f(A)= \frac{f(A_1)-f(A_2)}{A_1-A_2}\delta A \,,
\label{eq:dfA}
\end{equation}
for an arbitrary variation of $A$ (the labels 1 and 2 refer to $A$
before and after $\delta A$, respectively, following Convention 3). In
particular, $df(A)= (f(A_1)-f(A_2))/(A_1-A_2)dA$. Applying the
operator $d$ to $X$ as defined in \eq{v1}, and using the previous rule
to variate each of the arguments $A_i$ in $X$, immediately yields
\begin{equation}
dX = \Delta f(A_1,\dots,A_{n+1})(dA)^{n+1}\,,
\label{eq:n4a}
\end{equation}
with
\begin{equation}
\Delta f(z_1,\dots,z_{n+1})=
\sum_{k=1}^{n}
(-1)^{k+1} \frac{f(z_1,\dots,z_k,z_{k+2},\dots,z_{n+1})-
f(z_1,\dots,z_{k-1},z_{k+1},\dots,z_{n+1})}{z_k-z_{k+1}} \,.
\label{eq:n4}
\end{equation}
Because the operator $\Delta$ is a representation of the operator $d$
acting in the space of ordinary functions, it follows that
$\Delta^2=0$, as it is readily verified. The same operator appears
when the covariant derivative $\sDa$ is used, instead of $d$, although
in this case terms involving the field strength tensor $\sF$ are also
generated.

\subsection{Application to the Wess-Zumino-Witten action}
\label{subsec:II.B}

An interesting illustration of the usefulness of the Convention 3 can
be given by using it to explicitly integrate the WZW action
\cite{Witten:1983tw}. In two spacetime dimensions, the WZW functional
takes the form
\begin{equation}
\Gamma_{\text{WZW}}[U]=
 \frac{\eta_2}{4\pi}\int_{B_3} \Omega_3\,,\quad
\Omega_3=  -\frac{1}{3}\tr\left[(U^{-1}dU)^3\right]
 \,.
\end{equation}
The integration takes place in the interior of a three-dimensional
ball with a sphere S$^2$ (the compactified space-time) as
boundary. The field $U(x,t)$, which takes values on some matrix group,
interpolates between $u(x)$, at $t=1$ and a single point, say $U=1$,
at $t=0$. Because $\Omega_3$ is a closed 3-form and $u(x)$
contractile, the functional can be written as the integral of a 2-form
over S$^2$:
\begin{equation}
\Gamma_{\text{WZW}}[U]= 
\frac{\eta_2}{4\pi}\int_{\text{S}^2}\Omega_2
\end{equation}
with
\begin{equation}
\Omega_2=
-\int_0^1 dt\,\tr\left[(U^{-1}\partial_t U)(U^{-1}dU)^2\right] \,,
\end{equation}
and the result does not depend on the concrete interpolation.  We will
make use of our Convention 3 in order to explicitly carry out the
integration on the parameter $t$.  As interpolating field, let us
take\footnote{The choice $U(x,t)= 1+t(u(x)-1)$ is even simpler and, of
course, gives the same result, however, it may be disturbing that it
does not lie on the group manifold when $u(x)$ belongs to a group of
unitary matrices.}  $U(x,t)= u(x)^t= \exp({t\log u(x)})$. The branch
of the logarithm can be chosen with continuity because $u(x)$ is
contractile to 1, by assumption. Using the Convention 3
\begin{equation}
U^{-1}\partial_t U=\log u \,,\qquad
U^{-1}d U= \frac{1-(u_2/u_1)^t}{u_1-u_2}du\,,
\end{equation}
where the labels 1 and 2 refer to before and after $du$,
respectively. When these formulas are inserted in the expression of
$\Omega_2$, $U^{-1}\partial_t U$ carries a position label 1, the first
$U^{-1}d U$ block gives rise to labels 1 and 2, and second block to
labels 2 and 3. Due to the cyclic property $u_3$ is then identified
with $u_1$.\footnote{No confusion should arise with our previous
observation (cf. \eq{10}) that $\sm$ changes sign under the cyclic
property in the case of $W^-$, since the Conventions 1 and 2 are not
being used here. On the other hand, $du$ does change sign in
$\Omega_2$ under the cyclic property since it is a one-form.} This
gives
\begin{equation}
\Omega_2=
-\int_0^1 dt\,\tr\left[\log u_1 
\frac{1-(u_2/u_1)^t}{u_1-u_2}
\frac{1-(u_1/u_2)^t}{u_2-u_1}du^2
\right] \,.
\end{equation}
The point of following this procedure is that the dependence on $t$ is
now explicit and $u_1$ and $u_2$ are effectively c-numbers, therefore
the integration over $t$ is immediate
\begin{eqnarray}
\Omega_2 &=& \tr\left[h_{\text{WZW}}(u_1,u_2)du^2 \right] \,,
\label{eq:v3}
\end{eqnarray}
where the function $h_{\text{WZW}}$ is given by
\begin{eqnarray}
h_{\text{WZW}}(z_1,z_2) &=& \frac{1}{z_1-z_2}\left(
\frac{\log(z_1)-\log(z_2)}{z_1-z_2}-\frac{1}{2}\left(
\frac{1}{z_1}+\frac{1}{z_2}\right) \right) \,.
\label{eq:n25f}
\end{eqnarray}
It should be noted that, due to the cyclic property, the relation in
\eq{v3} does not uniquely determine $h_{\text{WZW}}(z_1,z_2)$ unless
the further constraint
$h_{\text{WZW}}(z_1,z_2)=-h_{\text{WZW}}(z_2,z_1)$ is imposed. On top
to this, a symmetric component can be added which does not contribute
inside the trace. In actual applications of the formula the purely
antisymmetric version of $h_{\text{WZW}}$ is clearly preferred since
an unsymmetrized function (although not $\Omega_2$ itself) could in
general present spurious singularities at $u_1=u_2$ as well as
spurious scale violations. The latter refers to the
following. $\Omega_3$ is invariant under an arbitrary local rescaling
of $U$, $U(x,t)\to \lambda(x,t)U(x,t)$, where $\lambda$ is a
c-number. Because $\Omega_2$ is unique in some sense to be discussed
below, it must also display this invariance.  The invariance under a
global rescaling already implies that (the symmetrized version of)
$h_{\text{WZW}}$ must be an homogeneous function; the possible
breaking introduced by the logarithm is canceled in this
version. Further, the invariance under a local rescaling is also
preserved due to $h_{\text{WZW}}(z,z)=0$ in the antisymmetric version.

The precise statement is that $h_{\text{WZW}}$ is the unique function
that works for a generic gauge group, that is, if no further
assumptions are made on the algebraic properties of the field
$u(x)$. For instance, any function $h$ would give the correct
vanishing result in the particular case of an Abelian gauge
group. Another interesting case is that of $u\in{\rm SU(2)}$. For this
group $u+u^{-1}$ is c-number and the same goes for any function $g(u)$
such that $g(z)=g(z^{-1})$. This is sufficient to show that, for any
antisymmetric function $h(z_1,z_2)$,
\begin{eqnarray}
\tr\left[h(u_1,u_2)(u^{-1}du)^2\right]
=\tr\left[h(u,u^{-1})(u^{-1}du)^2\right]\,,\quad(u\in{\rm SU(2)})\,,
\end{eqnarray}
thus, in particular
\begin{eqnarray}
\Omega_2 &=& \tr\left[h_{\text{WZW}}(u)(u^{-1}du)^2 \right] \,,
\quad(u\in{\rm SU(2)})\,,
\label{eq:v3b}
\end{eqnarray}
with
\begin{eqnarray}
h_{\text{WZW}}(z) &=&
\frac{4\log(z)-z^2+z^{-2}}{2(z-z^{-1})^2}\,.
\label{eq:n25g}
\end{eqnarray}

It is interesting to note that, in principle, the function
$h_{\text{WZW}}(z_1,z_2)$ can also be determined through an equation
involving only ordinary functions and no differential forms. This
comes about as follows.  The equation to be solved is
$\Omega_3=d\Omega_2$, where $\Omega_2$ is the unknown. For the latter,
the general form in \eq{v3} is proposed, whereas $\Omega_3$ can be
rewritten as
\begin{equation}
\Omega_3= \tr\left[-\frac{1}{3}\frac{1}{U_1U_2U_3}dU^3 \right] \,.
\end{equation}
The relation in \eq{n4a} then implies
\begin{equation}
-\frac{1}{3}\frac{1}{z_1z_2z_3}=
\frac{1}{3}\left(\Delta h_{\text{WZW}}(z_1,z_2,z_3)+
\Delta h_{\text{WZW}}(z_2,z_3,z_1)+\Delta h_{\text{WZW}}(z_3,z_1,z_2)\right)\,,
\label{eq:v5}
\end{equation}
where the cyclic property has been used to be able to equate both
sides of the equation.\footnote{It should be noted that the operator
$\Delta$ does not commute with the operation $P$ of projecting the
component which is invariant under cyclic permutations. Thus if
$\Delta$ is now applied to the right-hand side of \eq{v5} the result
does not vanish (despite the property $\Delta^2=0$) but it does vanish
after a subsequent application of $P$. This expresses the fact that
the 3-form $\Omega_3$ is closed.} Because of the lack of the
appropriate mathematical techniques, this kind of equation does not
seem to be particularly useful to determine the function $h_{\text{WZW}}$,
nevertheless it has the merit of reducing a problem of differential
forms to one of ordinary functions. Certainly it serves to check our
previous result for $h_{\text{WZW}}$.

The analogous expressions in four dimensions are
\begin{equation}
\Gamma_{\text{WZW}}[U]=
 \frac{\eta_4}{48\pi^2}\int_{B_5} \Omega_5\,,\quad
\Omega_5=  -\frac{1}{5}\tr\left[(U^{-1}dU)^5\right]
 \,.
\end{equation}
\begin{eqnarray}
\Omega_4 &=& \tr\left[\left(
\frac{1}{u_{12}u_{23}u_{34}u_{41}}
\left(
\frac{u_1}{u_2}
+\frac{u_1}{u_3}
+\frac{u_1}{u_4}
-\frac{1}{2}\frac{u_1u_3}{u_2u_4}
\right)
+2\frac{u_{12}-u_{41}}{u_{12}^2 u_{13} u_{41}^2}
\log(u_1)
\right)
du^4
\right] \,,
\end{eqnarray}
where $u_{ij}=u_i-u_j$. For the sake of shortness the function has
not been explicitly symmetrized in order to extract its invariant
component under cyclic permutations. As noted before in the two
dimensional case, such a symmetrization is needed in practice.

Because in the four-dimensional case the integral refers to a 4-form,
the formula seems to predict a vanishing value (or more generally, a
multiple of $2\pi i$ ) for the WZW term when the gauge group is
three-dimensional such as SU(2), whereas actually the result is a
multiple of $i\pi$. However, the $i\pi$ result corresponds to
configurations which cannot be contracted within SU(2). Another
observation is that the use of arbitrary functions $h(z_1,z_2)$ in two
dimensions, or $h(z_1,z_2,z_3,z_4)$ in four dimensions, allows to
propose phenomenological contributions to the effective action in the
pseudo-parity odd sector, which are more general that the usual WZW
term. All these possible new contributions are automatically invariant
under global vector transformations ($u\mapsto\Omega^{-1}u\Omega$ with
constant $\Omega$). Among them, the WZW term is singularized because
it is invariant under global chiral transformations
($u\mapsto\Omega_L^{-1}u\Omega_R$ with constant $\Omega_{L,R}$). On
the other hand, it can be noted that $\Omega_2$ (or $\Omega_4$ in four
dimensions) is not the unique solution of $\Omega_3=d\Omega_2$, since
$\Omega_2+d\omega$ ($\omega$ being an arbitrary 1-form) would also be
a solution. $\Omega_2$ is singularized because it is the one solution
which is manifestly invariant under global vector transformations.

\section{Explicit computation of the covariant current}
\label{sec:4}

As stated above, our purpose is to compute the leading term of the
pseudo-parity odd component of the effective action of Dirac
fermions. By leading term we mean that with the less number of
covariant derivatives and covariant will always refer to chiral gauge
transformations. Because $W^-[v,m]$ contains the Levi-Civita
pseudo-tensor, the leading term is that with $d$ Lorentz indices, $d$
being the space-time dimension, which is assumed to be even. In
practice we will consider $d=0,2,4$. All other higher order terms in
the derivative expansion are ultraviolet finite and thus free from
anomalies and multivaluation. The chiral anomaly, multivaluation and
topological pieces of the effective action are contained in the
leading term. There are no other anomalies (such as scale or parity
anomalies) in $W^-[v,m]$ in even dimensions.  Since no higher orders
will be considered in this work, from now on $W^-[v,m]$ will be used
to refer to the leading term. We will always work with the LR version
of the effective action except in Section \ref{subsec:V.f} and
Appendix \ref{app:A}.

\subsection{The covariant current}

Due to the presence of the chiral anomaly in the pseudo-parity odd
component of the effective action, $W^-[v,m]$ is not a chiral
invariant functional and this makes advisable to use an indirect
procedure to compute it. We will adopt the traditional Schwinger's
approach \cite{Schwinger:1951nm} of working with the current, i.e. the
variation of the effective
action~\cite{Leutwyler:1985em,Ball:1989xg}. The reason of course is
that there is a version of the current which is chiral covariant and
thus easier to treat.

For subsequent reference, we note that there are two quantities to be
distinguished: the effective ``current'' $\sJ_v^-$ which is related to
the variation with respect the gauge fields, and the effective
``density'' $\sJ_m^-$ which is the variation with respect to the
scalar fields. The consistent effective current and density will be
defined as
\begin{equation}
\delta W^-[v,m]= \left\langle
\sJ_v^-\delta\sv+\sJ_m^-\delta\sm \right\rangle\,.
\label{eq:curr}
\end{equation}
Our conventions 1 and 2 are being used, $\delta\sm$ and $\delta\sv$
are arbitrary variations of the external fields, and $\delta\sm$,
$\delta\sv$, $\sJ_v^-$ and $\sJ_m^-$ are $0$-, $1$-, $(d-1)$- and
$d$-forms, respectively. $\langle\ \rangle$ was defined in \eq{n2}.

In addition, one has to distinguish between the consistent and the
covariant currents. The former is the variation of the effective
action but it fails to be chiral covariant due to the presence of the
chiral anomaly. On the other hand, the chiral covariant version
$\sJ_{v,c}^-$ is not consistent, i.e., is not a true variation.
Both versions of the effective current
are realizations of the same formal object. This means that they
coincide in their ultraviolet finite pieces and so they differ only by
a counterterm which is a polynomial in the external fields and their
derivatives:
\begin{equation}
\sJ_v^- = \sJ^-_{v,c} +\sP(v) \,.
\label{eq:v10}
\end{equation}
$\sP(v)$ is a fixed known polynomial which depends solely on the gauge
fields and its derivatives. This polynomial is purely geometrical in
the same sense as the chiral anomaly, and in fact is is completely
determined by the anomaly \cite{Bardeen:1984pm}.

The idea of the calculation of $W^-[v,m]$ is as follows. We will
explicitly compute the covariant current, then we will write the most
general form $W^-[v,m]$ consistent with chiral and Lorentz symmetries,
with some functions as unknowns, and finally these unknowns will be
chosen so as to reproduce the current. It is only necessary to make
sure that the effective action is uniquely determined by this
procedure. That this will be the case can be seen by the following
argument. Let $A^-[v,m]$ denote a possible ambiguity in the effective
action allowed by this procedure.  Because the current is reproduced,
$A^-[v,m]$ must actually be a functional of $m$ only. In addition,
$A^-[m]$ must be chiral invariant, since we have already imposed the
correct chiral transformation on our functional.  Then it can be
evaluated in any chirally rotated configuration, and in particular one
can always choose $\mlr=\mrl$. It follows that the ambiguity vanishes
since this functional is odd under pseudo-parity, i.e., under exchange
of the labels $L$ and $R$.

It is also possible to use the density $\sJ^-_m$ instead of the
current. In this case the ambiguity can only be a function of $v$ and
it is easily shown that no such chiral invariant functional exists, at
leading order. (The previous argument for the current holds, however,
to all orders in the derivative expansion.) An advantage of $\sJ^-_m$
would be that there is no distinction between consistent and covariant
density (the consistent density is automatically chiral covariant).
Nevertheless, within a derivative expansion, the current is preferable
for purely technical reasons, namely, the current contains $d-1$
derivatives whereas the density contains $d$ and thus it requires more
work. Explicit formulas for the effective density in two and four
dimensions are given in Section \ref{subsec:V.e}.

In order to highlight the main results of this section, the
calculation itself will be deferred until the end of this section. As
will be clear from the calculation below, the general form of the
covariant current in two and four dimensions (of course, at leading
order in the derivative expansion) is
\begin{eqnarray}
\sJ^-_{v,c,d=2}
 &=& A(\sm_1,\sm_2)\sm'
\,, \nonumber \\
\sJ^-_{v,c,d=4} &=& 
A(\sm_1,\sm_2,\sm_3,\sm_4)\sm'^3 + A(\sm_1,\sm_2,\sm_3,)\sF\sm'
+ A'(\sm_1,\sm_2,\sm_3)\sm'\sF \,
\label{eq:v7}
\end{eqnarray}
where $\sm'$ denotes the 1-form $\sDa\sm$. The subindex $c$ in the
currents recalls that this is the covariant current. The various
symbols $A$ denote different known functions. We will often use the
shorthand notation $A_{12}$ to denote $A(\sm_1,\sm_2)$, $A_{123}$ to
denote $A(\sm_1,\sm_2,\sm_3)$, etc. In addition, $A_{\m{1}2}$ will
denote $A(-\sm_1,\sm_2)$, etc.

As we have just mentioned, the formulas in \eq{v7} follow from the
explicit calculation, nevertheless, by now it is probably already
obvious that they are just the most general possible form for the
currents at leading order consistent with Lorentz and chiral gauge
invariance. Let us see which properties are to be expected for the
functions $A$ in $\sJ^-_{v,c}$. Because there is no scale anomaly in
$W^-[v,m]$, $A$ should homogeneous functions of the appropriate
degree. Next, there is the {\em consistency condition} that in each
term of $\sJ_{v,c}^-$ there should be as many $L$ labels as $R$
labels, thus $\sm$ must appear an even number of times
\begin{equation}
A_{12}=-A_{\m{1}\m{2}}\,,\quad
A_{123}=-A_{\m{1}\m{2}\m{3}}\,,\quad
A'_{123}=-A'_{\m{1}\m{2}\m{3}}\,,\quad
A_{1234}=-A_{\m{1}\m{2}\m{3}\m{4}}\,,
\end{equation}
(that is $A(\sm_1,\sm_2)= A(-\sm_1,-\sm_2)$, etc).

A further condition is implied by the fact that $W^-[v,m]$ is purely
imaginary. First note that the functions $A$ are all purely real since
there are no $i$'s in the formulas nor can they be generated during
the calculation, except through $\gamma_5$ when $d=4n+2$. The possible
factor $i$ is explicit through $\eta_d$ in the normalization of
$\langle\ \rangle$ (cf. \eq{n2}). On the other hand, the fact that all
quantities involved behave in a well defined way under Hermitian
conjugation allows to reformulate this conjugation in terms of an
equivalent {\em mirror transformation} which has the advantage of
being purely algebraic (no complex conjugation is involved).  Such a
mirror transformation is defined by the following rules i) the
elementary objects $\sm$, $\sv$ (or $\delta\sv$) and $\sD$ are mirror
invariant, ii) the transformation is linear, and iii) the order of the
factors is transposed (regardless of whether they are functions or
differential forms). The transformation of derived quantities follows
from the previous rules, thus $\sF\to \sF$, $d\sm\to -d\sm$, $d\sv\to
d\sv$, $\sDa\to\mp\sDa$ (depending on whether it acts commuting or
anticommuting, respectively), $\sm'\to -\sm'$, etc. For instance
\begin{eqnarray}
&& \langle\sm\sv\sm^{-1}\sv\rangle \to
\langle\sv\sm^{-1}\sv\sm\rangle = 
-\langle\sm\sv\sm^{-1}\sv\rangle \,,
\nonumber \\
&&
A(\sm_1,\sm_2)\sm' \to -A(\sm_2,\sm_1)\sm'
\,.
\end{eqnarray}
The antihermiticity of $W^-[v,m]$ implies that this quantity is odd
under the mirror transformation whereas $\sJ^-_v$ and $\sJ_m^-$ are
even. Therefore the following conditions are found
\begin{equation}
A_{12}=-A_{21}\,,\quad
A'_{123}=-A_{321}\,, \quad
A_{1234}=-A_{4321}\,.
\end{equation}
Thus the function $A'_{123}$ is not independent.

Finally, there is an extremely important property satisfied by these
functions which is finiteness in the coincidence limit. This refers to
the following.  The most general form of $A_{12}$ allowed by
consistency and mirror symmetry, is
\begin{equation}
A_{12}= \sm_1 f(\sm_1^2,\sm_2^2)-\sm_2 f(\sm_2^2,\sm_1^2)\,,
\end{equation}
for certain function $f$. Inserting this general form in the
expression of $\sJ^-_{c,v}$ in \eq{v7}, and making explicit the chiral
labels, yields
\begin{equation}
(\sJ^-_{c,v})_R = 
f(\sm_{R1}^2,\sm_{R2}^2) (\sm\sm')_R
- f(\sm_{R1}^2,\sm_{R2}^2) (\sm'\sm)_R \,.
\end{equation}
As noted before, in order to numerically evaluate this expression, a
natural procedure is to use a basis of eigenvectors of $\sm^2$
\cite{Salcedo:2000I}. In this way, $(\sm\sm')_R$ and $(\sm'\sm)_R$ are
replaced by matrix elements whereas the $\sm^2$ are replaced by
eigenvalues. In particular, in the diagonal matrix elements, $\sm_1^2$
and $\sm^2_2$ take the same value (note that $\mlr\mrl$ and $\mrl\mlr$
are related by a similarity transformation and thus they have the same
eigenvalues). Finiteness of the current requires that $f$ must be
finite as its two arguments coincide. (Because the terms with
$\sm\sm'$ and $\sm'\sm$ have different chiral labels no cancellation
can take place among them in general.)  In summary, the functions
$A_{12}$, $A_{123}$, etc, must be regular as two or more arguments
coincide up to a sign. This is automatically satisfied by the true
functional describing the current, as no physical singularity exists
in the coincidence limit (cf. eqs. (\ref{eq:n10}) and (\ref{eq:n10a})
below) however, the formalism allows to write functionals which
violate this condition. Such functionals are only formal and are
meaningless or at least ambiguous. On the other hand, physical
singularities can occur as $\sm\to 0$ and they will reflected in the
effective action and currents.

After this discussion, let us quote the result coming from the
explicit calculation in two dimension
\begin{eqnarray}
A_{12} &=&
-\frac{2}{\sm_1-\sm_2} 
+\frac{2\sm_1\sm_2}
{(\sm_1-\sm_2)(\sm_1^2-\sm_2^2)}\log(\sm_1^2/\sm_2^2)
\,.
\label{eq:69a}
\end{eqnarray}
The corresponding four-dimensional formulas are collected in Appendix
\ref{app:B}. It can be checked that the full functions have all the expected
properties and in particular they preserve scale invariance and are
regular in the coincidence limits. At this point, it is perhaps worth
noticing another essential property of these functions, namely, they
are unambiguous. The analogous functions for the effective action are
not unique due to integration by parts and the trace cyclic
property. This not the case for the current; these functions are the
unique result of the calculation. The formulas can only be simplified
by considering particular cases, i.e., particular flavor groups.

Analyzing the form of the functions $A$ we have found it convenient to
introduce the auxiliary functions $\bar{A}$:
\begin{equation}
A_{12}=\bar{A}_{1\m{2}}\,,\quad A_{123}=\bar{A}_{12\m{3}}\,,\quad
A_{1234}=\bar{A}_{1\m{2}3\m{4}}\,,
\end{equation}
(i.e. $A(\sm_1,\sm_2)=\bar{A}(\sm_1,-\sm_2)$, etc). For these
functions, consistency and mirror symmetry translates into
\begin{equation}
\bar{A}_{12}=-\bar{A}_{\m{1}\m{2}}\,,\quad
\bar{A}_{123}=-\bar{A}_{\m{1}\m{2}\m{3}}\,,\quad
\bar{A}_{1234}=-\bar{A}_{\m{1}\m{2}\m{3}\m{4}}\,,\quad
 \bar{A}_{12}=\bar{A}_{21}\,,\quad \bar{A}_{1234}=\bar{A}_{4321}\,. 
\end{equation}
As we will see later, further conditions are implied by the fact that
the underlying theory is Lorentz and chiral covariant. In particular,
this implies
\begin{equation}
\bar{A}_{123}= \bar{A}_{132}\,.
\end{equation}
Remarkably, the true functions $\bar{A}$, i.e. those resulting from
the calculation, in four dimensions turn out to have a larger
symmetry, namely, they are completely symmetric functions of their
arguments:
\begin{equation}
\bar{A}_{12}=\bar{A}_{21}\,,\quad
\bar{A}_{123}=\bar{A}_{213}=\bar{A}_{231} \,,\quad
\bar{A}_{1234}= \bar{A}_{2134}= \bar{A}_{2341}\,. 
\end{equation}
(The complete symmetry also holds in two dimensions but it this case
this follows from previous symmetries.) It is not clear why, in the
four dimensional case, the symmetry is larger than expected. This
symmetry does not follows from Lorentz invariance and (anomalous)
chiral symmetry since it is possible to write Lorentz invariant
functionals with the correct chiral anomaly but with associated
variations which are not symmetric functions under permutation of
their arguments (see \eq{n12} below). It seems to be a property of the
true current only. The same symmetry is also found for the effective
density in two and four dimensions (see Section \ref{subsec:V.e}).

\subsection{Explicit computation of the covariant current}
\label{subsec:III.b}

Let us consider a first order variation of the effective action. This
is formally given by
\begin{equation}
\delta W[v,m] = -\Tr\left(\frac{1}{\Dirac}\delta\Dirac\right)\,.
\end{equation}
The variation of the Dirac operator is
\begin{equation}
\delta\Dirac= 
  P_R\delta\mlr P_R
+ P_R\,\delta\!\thru{v}_LP_L
+ P_L\delta\!\thru{v}_R P_R 
+ P_L\delta\mrl P_L \,.
\end{equation}
On the other hand, the inverse Dirac operator can be written
as\footnote{This follows from writing the Dirac operator as
\begin{equation}
\Dirac = \left(\matrix{ \mlr & \thru\D_L \cr \thru\D_R & \mrl }\right)\,,
\end{equation}
(where actually the $\gamma_\mu$ stand for submatrices of half
dimension after restriction to the $LR$ of $RL$ sectors) and then
using the matrix identity
\begin{equation}
\left(\matrix{ A & B \cr C & D }\right)^{-1} =
\left(\matrix{ (A-BD^{-1}C)^{-1} &  (C-DB^{-1}A)^{-1} \cr
(B-AC^{-1}D)^{-1} & 
(D-CA^{-1}B)^{-1} }\right),
\end{equation}
where $A$, $B$, $C$, $D$ are square submatrices. This formula can be
rewritten in a way that holds too when $A$ and $D$ have different
dimension and thus $B$ and $C$ are not square matrices.}
\begin{eqnarray}
\Dirac^{-1} &=& 
P_R(\mlr-\thru\D_L\mrl^{-1}\thru\D_R)^{-1}P_R +
P_R(\thru\D_R-\mrl\thru\D_L^{-1}\mlr)^{-1}P_L
\nonumber \\ &&  +
P_L(\thru\D_L-\mlr\thru\D_R^{-1}\mrl)^{-1}P_R +
P_L(\mrl-\thru\D_R\mlr^{-1}\thru\D_L)^{-1}P_L \,.
\end{eqnarray}
Therefore the variation of the effective action is
\begin{eqnarray}
\delta W[v,m] &=& -\Tr\Bigg[
P_R(\mlr-\thru\D_L\mrl^{-1}\thru\D_R)^{-1} \delta\mlr +
P_R(\thru\D_R-\mrl\thru\D_L^{-1}\mlr)^{-1} \delta\thru{v}_R
\nonumber \\ && 
 + P_L(\mrl-\thru\D_R\mlr^{-1}\thru\D_L)^{-1}\delta\mrl
 + P_L(\thru\D_L-\mlr\thru\D_R^{-1}\mrl)^{-1}\delta\thru{v}_L
\Bigg]\,.
\label{eq:n6a}
\end{eqnarray}
This variation can be separated into its pseudo-parity even (without
$\gamma_5$) and odd (with $\gamma_5$) components. Then, Conventions 1
and 2 can directly be applied and this yields
\begin{eqnarray}
\delta W^+[v,m] &=& -\Tr\left[
(\sm -\thru\sD\sm^{-1}\thru\sD)^{-1} \delta\sm +
(\thru\sD-\sm\thru\sD^{-1}\sm)^{-1} \delta\thru{\sv} \right]\,,
\nonumber \\
\delta W^-[v,m] &=& -\Tr\left[ \gamma_5\left(
(\sm -\thru\sD\sm^{-1}\thru\sD)^{-1} \delta\sm +
(\thru\sD-\sm\thru\sD^{-1}\sm)^{-1} \delta\thru{\sv}\right) \right]\,.
\end{eqnarray}
Once our conventions are used, the variations can be rewritten in the
simpler form
\begin{eqnarray}
\delta W^+[v,m] &=& -\Tr\left[
\frac{1}{\thru\sD+\sm}(\delta\!\thru{\sv}+\delta\sm)
\right]\,,
\nonumber \\ 
\delta W^-[v,m] &=& -\Tr\left[\gamma_5
\frac{1}{\thru\sD+\sm}(\delta\!\thru{\sv}+\delta\sm)
\right]\,.
\label{eq:n6}
\end{eqnarray}
(Actually, what enters is
$$\frac{1}{2}\left(
(\thru\sD+\sm)^{-1}(\delta\!\thru{\sv}+\delta\sm)
+ (\thru\sD-\sm)^{-1}(\delta\!\thru{\sv}-\delta\sm)
\right)\,,$$
but the even component under $\sm\to -\sm$ is automatically selected
by the Dirac trace.)

The variation of the pseudo-parity even component is just
\begin{equation}
\delta W^+[v,m] = -\delta\Tr\log[\thru\sD+\sm]\,.
\end{equation}
Therefore, within our notation, $W^+[v,m]$ is completely identical to
a purely vector-like theory (i.e., one with $v_R=v_L$ and
$\mlr=\mrl$), a fact already exploited in \cite{Salcedo:2000I}.

The pseudo-parity odd case is different. $\delta W^-[v,m]$ cannot be
expressed as the variation of a functional of the form
$\Tr[\gamma_5f(\sm,\sD)]$, since that would not allow for the chiral
anomaly. Technically the difference with the pseudo-parity even case
comes from the cyclic property which is affected by the presence of
$\gamma_5$ as well as by the different behavior of $\sm$,
cf. \eq{10}. Thus there is an obstruction to integrate the variation
preserving all symmetries\cite{Leutwyler:1985em,Ball:1989xg}. No such
problem arises if one wants to compute just the current or the
density: because one particular operator is distinguished, namely
$\delta\sv$ or $\delta\sm$, the cyclic property is no longer required
and the anomalous behavior of $\sm$ under the cyclic property does not
enter.

Comparing with its definition in  \eq{curr}, the current can be
formally read off from
\begin{eqnarray}
\delta W^-[v,m] &=& \int d^dx\,\tr\left[\delta\sv_\mu \langle
x|\gamma_5\gamma_\mu \frac{1}{\thru\sD+\sm} |x \rangle \right]
\,,\quad (\delta\sm=0)
\label{eq:curr1}
\end{eqnarray}
where the trace includes flavor and Dirac spaces. This is formal
because the matrix element in the right-hand side is ultraviolet
divergent and needs to be given a meaning through some renormalization
procedure. Noting that the cyclic property does not enter and in
addition no $\gamma_5$ appears in $(\thru\sD+\sm)^{-1}$, it follows
that the symbols $\sD$ and $\sm$ behave algebraically as those of an
effective vector-like theory. This allows to use a regularization
prescription preserving the corresponding vector gauge invariance.
Such effective vector gauge invariance amounts to chiral covariance
for the operator $(\thru\sD+\sm)^{-1}$ and therefore this procedure
will yield the chiral covariant effective current.

In the particular case of two spacetime dimensions, there is a
shortcut. The two dimensional identity
$\gamma_5\gamma_\mu=-\eta_2\epsilon_{\mu\nu}\gamma_\nu$ allows to relate
$\delta W^-[v,m]$ with a variation of $W^+[v,m]$, namely
\begin{eqnarray}
\delta W^-[v,m] &=& 
\int d^2x\,\tr\left[\eta_2\epsilon_{\mu\nu}\delta\sv_\mu
\frac{\delta W^+}{\delta\sv_\nu}
\right] \,.
\end{eqnarray}
Use of the result in \cite{Salcedo:2000I},
\begin{equation}
W^+_{2,2}[v,m] = -\frac{1}{4\pi}\int d^2x\,\tr\left[
\left(\sm_1\sm_2\frac{\log(\sm_1^2/\sm_2^2)}{\sm_1^2-\sm_2^2}
-1\right) \frac{(\sDa_\mu\sm)^2}{(\sm_1-\sm_2)^2}
  \right]
\label{eq:13}
\end{equation}
directly produces
\begin{equation}
\sJ^-_{v,c} = 
\frac{2}{\sm_1-\sm_2}\left(\sm_1\sm_2 
\frac{\log(\sm_1^2/\sm_2^2)}{\sm_1^2-\sm_2^2}-1\right)\sDa\sm \,,
\label{eq:n8}
\end{equation}
where the labels $1$ and $2$ refer to {\em before} and {\em after}
$\sDa\sm$ respectively. From this formula, one can immediately read
off the function $A_{12}$ introduced in \eq{v7}, and this gives the
result quoted in \eq{69a}. In the two dimensional case, there is yet
another method which yields the effective action directly from the
anomaly. This method is explained in Section \ref{subsec:V.d}.

In order to compute $\sJ^-_{v,c}$ beyond two dimensions we will use
the convenient method introduced by Pletnev and Banin
~\cite{Pletnev:1999yu}. The method can be briefly summarized as
follows: Let $f(\sm,\sD)$ be an operator constructed out of $\sm$ and
$\sD_\mu$. In the usual symbols method (see
e.g. \cite{Salcedo:1996qy})
\begin{equation}
\langle x|f(\sm,\sD)|x\rangle= 
\int\frac{d^dp}{(2\pi)^d}
\langle x|f(\sm,\sD+p)| 0\rangle \,,
\end{equation}
where $|0\rangle$ is the state with zero wavenumber, i.e. $\langle
x|0\rangle=1$, and the momentum $p_\mu$ is just a
c-number.\footnote{Our notation will be as follows: $p_\mu$ is purely
imaginary, however, $\int d^dp$ denotes the standard integration on
${\mathbb{R}}^d$ and $p^2$ denotes $-p_\mu p_\mu$.}  The matrix
element $\langle x|f(\sm,\sD)|x\rangle$ is manifestly gauge covariant,
however $\langle x|f(\sm,\sD+p)| 0\rangle$ is not, because of
$|0\rangle$. Gauge invariance is recovered only after momentum
integration. This nuisance is avoided by Pletnev and Banin by
considering
\begin{eqnarray}
\langle x|f(\sm,\sD)|x\rangle &=& 
\int\frac{d^dp}{(2\pi)^d}
\langle x|\exp(-\partial_p\sD) f(\sm,\sD+p)\exp(\partial_p\sD)|0\rangle \,,
\nonumber \\ &=&
\int\frac{d^dp}{(2\pi)^d}
\langle x| f(\bar{\sm},\bar{\sD})| 0\rangle \,,
\end{eqnarray}
The first equality follows because the momentum derivative
$\partial^p_\mu=\partial/\partial p_\mu$ in the last
$\exp(\partial_p\sD)$ factor has no effect since there are no $p_\mu$
dependence at its right. Similarly the first factor
$\exp(-\partial_p\sD)$ changes nothing, by integration by parts in the
momentum integration. The second equality uses that
$\exp(-\partial_p\sD)\sX\exp(\partial_p\sD)$ defines a similarity
transformation.\footnote{Actually the full similarity transformations
is
$$\sX\to\bar{\sX}=
\exp(-\partial_p\sD)\exp(-xp)\sX\exp(xp)\exp(\partial_p\sD)\,.$$ The
inner transformation produces $\sm\to\sm$ and
$\sD_\mu\to\sD_\mu+p_\mu$, and is the one used to arrive to the
symbols method formula.} Explicit computation gives \cite{Pletnev:1999yu}
\begin{eqnarray}
{\bar{\sm}} &=& \sm -\sDa_\mu\sm\,\partial^p_\mu
+\frac{1}{2!}\sDa_\nu\sDa_\mu\sm\,\partial^p_\nu\partial^p_\mu
-\frac{1}{3!}\sDa_\alpha\sDa_\nu\sDa_\mu\sm\,\partial^p_\alpha\partial^p_\nu\partial^p_\mu
+\cdots \,, \nonumber \\ {\bar{\sD_\mu}} &=& p_\mu
-\frac{1}{2!}\sF_{\nu\mu}\,\partial^p_\nu
+\frac{2}{3!}\sDa_\alpha\sF_{\nu\mu}\,\partial^p_\alpha\partial^p_\nu
-\frac{3}{4!}\sDa_\beta\sDa_\alpha\sF_{\nu\mu}\,\partial^p_\beta\partial^p_\alpha\partial^p_\nu
+\cdots \,.
\label{eq:n7}
\end{eqnarray}
As usual $\sDa_\mu\sX$ stands for $[\sD_\mu,\sX]$, the chiral
covariant derivative of $\sX$. The operator $\partial^p_\mu$ denotes
the derivative with respect to the $p_\mu$ dependence. It acts
derivating everything to its right (or to its left, by parts).  The
point of doing this is that the operators $\partial_\mu$ (derivative
with respect to $x_\mu$) appear only through $\sDa_\mu$ and so i)
gauge covariance is manifest and ii) the integrand is just a function
of $x$ (rather than a pseudo-differential operator as
$f(\sm,\sD)$). This last fact allows to write
\begin{eqnarray}
\langle x|f(\sm,\sD)|x\rangle &=& 
\int\frac{d^dp}{(2\pi)^d} f(\bar{\sm},\bar{\sD}) \,,
\end{eqnarray}
where $f(\bar{\sm},\bar{\sD})$ is a matrix valued function of $x$.

In our case an application of this method amounts to replacing
\eq{curr1} by
\begin{eqnarray}
\delta W^-[v,m] &=& 
\int d^dx\,\tr\left[\gamma_5 \delta\!\thru{\sv}
\int\frac{d^dp}{(2\pi)^d}
\frac{1}{\thru{\bar{\sD}}+{\bar{\sm}}}
\right]
\,,
\label{eq:n5a}
\end{eqnarray}
Note that $\tr$ refers to Dirac and flavor spaces here.

The calculation proceeds as follows. The formula (\ref{eq:n5a}) is
expanded in the number of covariant derivatives, or equivalently in
the number of Lorentz indices carried by $\sDa_\mu$ and
$\sF_{\mu\nu}$. At leading order the term with $d-1$ spatial indices
is selected. The derivatives with respect to $p_\mu$ are carried
out. The Dirac trace is taken. This produces a Levi-Civita
pseudo-tensor and differential geometry notation can be used. Note
that terms with two or more $\partial^p_\mu$ in ${\bar{\sm}}$ and
${\bar{\sD_\mu}}$ cancel since the corresponding indices are
symmetrized. Next, the $\sm$ are indexed according to Convention 3
thereby becoming c-numbers. This allows to carry out the momentum
integrations straightforwardly; the integration formulas of
\cite{Salcedo:2000I} apply.

A technical detail is that, computationally, the Dirac algebra is
slightly alleviated by rewriting \eq{curr1} as
\begin{eqnarray}
\delta W^-[v,m] &=& 
\int d^dx\,\tr\left[
\langle x|\gamma_5\delta\!\thru{\sv}
\frac{1}{(\thru\sD-\sm)(\thru\sD+\sm)}
(\thru\sD-\sm) |x \rangle \right]
\nonumber \\
 &=& 
-\int d^dx\,\tr\left[
\langle x|\gamma_5\delta\!\thru{\sv}
\frac{1}{-\sD_\mu^2+\sm^2-\thru{\sDa}\!\sm
-\frac{1}{2}\sigma_{\mu\nu}\sF_{\mu\nu}}
(\thru\sD-\sm) |x \rangle \right]
\label{eq:n5b}
 \,.
\end{eqnarray}
The formulas in \eq{n7} define a similarity transformation
~\cite{Pletnev:1999yu} so the replacements $\sm\to\bar{\sm}$ and
$\sD_\mu\to\bar{\sD}_\mu$ apply here too.

Let us illustrate this procedure for the two-dimensional
case. Applying the replacements $\sm\to\bar{\sm}$ and
$\sD_\mu\to\bar{\sD}_\mu$ in the second \eq{n5b}, and retaining terms
with at most one covariant derivative, yields
\begin{eqnarray}
\delta W^-_{d=2}[v,m] &=& -\int\frac{d^2xd^2p}{(2\pi)^2}\tr\left[
\gamma_5\delta\!\thru{\sv}\frac{1}{\Delta-\thru{\sDa}\!\sm-\{\sm,\sDa_\mu\sm\}
\partial^p_\mu}(\thru{p}-\sm) \right] \nonumber 
\\ &=&
-\int\frac{d^2xd^2p}{(2\pi)^2}\tr\left[
\gamma_5\delta\!\thru{\sv}\frac{1}{\Delta}
\left(\thru{\sDa}\!\sm+\{\sm,\sDa_\mu\sm\}
\partial^p_\mu\right) \frac{1}{\Delta}(\thru{p}-\sm) \right] \,,
\end{eqnarray}
where we have defined $\Delta= p^2+\sm^2$ (not to be confused with the
operator $\Delta$ introduced in \eq{n4}). The formula is already
ultraviolet convergent without further renormalization. This was to be
expected since the chiral covariant current is unique and thus free
from ultraviolet ambiguities. Using the formulas
\begin{equation}
\partial^p_\mu\frac{1}{\Delta}= \frac{2p_\mu}{\Delta^2}\,,\quad
\gamma_5\gamma_\mu\gamma_\nu \to -\eta_2\epsilon_{\mu\nu}\,, \quad
p_\mu p_\nu \to -\frac{p^2}{2}\delta_{\mu\nu} \,,
\end{equation}
the expression becomes
\begin{eqnarray}
\delta W^-_{d=2}[v,m]
&=& -2\eta_2\int\frac{d^2p}{(2\pi)^2}\tr\left[
\delta\sv\left(
\frac{1}{\Delta}\sm'\frac{\sm}{\Delta}
-p^2\frac{1}{\Delta^2}\{\sm,\sm'\}\frac{1}{\Delta}
\right)
\right] 
\nonumber \\
&=& 2\eta_2\int\frac{d^2p}{(2\pi)^2}\tr\left[
\left(
\frac{\sm_2}{\Delta_1\Delta_2}
-p^2\frac{\sm_1+\sm_2}{\Delta_1^2\Delta_2}
\right)\sm'\delta\sv
\right] 
 \,,
\label{eq:n10}
\end{eqnarray}
where we are already using a notation of differential forms, and
$\sm'=\sDa\sm$. The trace no longer includes Dirac space. The
integration over momenta can be done using the formulas in
\cite{Salcedo:2000I} and the result in \eq{n8} follows.

The calculation in four dimensions is similar and yields
\begin{eqnarray}
\delta W^-_{d=4}[v,m] 
&=&  4\eta_4\int\frac{d^4p}{(2\pi)^4}\tr\Bigg[ 
\nonumber \\
&& \left(
\frac{\sm_3}{\Delta_1\Delta_2\Delta_3\Delta_4}
+\frac{p^2}{2}\left(
\frac{\sm_1-\sm_3}{\Delta_1^2\Delta_2\Delta_3\Delta_4}
-\frac{\sm_2+\sm_3}{\Delta_1\Delta_2^2\Delta_3\Delta_4}
-\frac{\sm_3+\sm_4}{\Delta_1\Delta_2\Delta_3\Delta_4^2}
             \right)\right)\sm'^3\delta\sv
\nonumber \\
&& +\left(
\frac{\sm_1}{\Delta_1\Delta_2\Delta_3}
-\frac{p^2}{2}\left(
\frac{\sm_1-\sm_2}{\Delta_1\Delta_2^2\Delta_3}
+\frac{\sm_1+\sm_3}{\Delta_1\Delta_2\Delta_3^2}
             \right)\right)\sF\sm'\delta\sv
\nonumber \\
&& + \left(
\frac{\sm_1}{\Delta_1\Delta_2\Delta_3}
-\frac{p^2}{2}\left(
\frac{\sm_1+\sm_2}{\Delta_1\Delta_2^2\Delta_3}
+\frac{\sm_1+\sm_3}{\Delta_1\Delta_2\Delta_3^2}
             \right)\right)\sm'\sF\delta\sv
\Bigg] 
 \,.
\label{eq:n10a}
\end{eqnarray}
Integration over momentum yields the results quoted in \eq{69b}.  It
can be noted that the integrands in eqs. (\ref{eq:n10}) and
(\ref{eq:n10a}) are not unique, due to integration by parts in
momentum space. On the other hand, their integral, the functions $A$,
are unambiguous.

\section{The effective action}
\label{sec:3}

Following the strategy outline above, we should now consider the most
(or, at least, a sufficiently) general effective action functional in
the pseudo-parity odd sector and at leading order in the covariant
derivative expansion, consistent with Lorentz and chiral symmetries.
This will be done by writing the effective action as
\begin{equation}
W^-[v,m]=  \Gamma_{\text{gWZW}}[v,m] + W^-_c[v,m]\,.
\label{eq:n36}
\end{equation}
The functional $\Gamma_{\text{gWZW}}[v,m]$, an extended gauged
Wess-Zumino-Witten (gWZW) action, is chosen in order to reproduce the
correct chiral anomaly. The extension refers to the fact that it goes
beyond the chiral circle constraint. Once the anomaly is saturated,
the remainder will be chiral invariant and can be adjusted in order to
reproduce the known current. This chiral invariant remainder is
denoted by  $W^-_c[v,m]$.

\subsection{The extended gauged Wess-Zumino-Witten action}
\label{subsec:IV.a}

As is well-known (see Appendix \ref{app:A}), the ordinary gauged WZW
functional $\Gamma_{\text{LR}}[v_L,v_R,U]$ reproduces the correct
chiral anomaly (in the LR version). Two essential properties of this
result are i) that it follows solely from assuming the transformation
property $U\to \Omega_L^{-1}U\Omega_R$, and no other algebraic
properties on $U(x)$, and ii) the infinitesimal chiral variation of
$\Gamma_{\text{LR}}[v_L,v_R,U]$ (i.e. the anomaly) depends on the
gauge fields $v_{L,R}$ but not on $U$.  In view of this, we can use
$\mlr$ instead of $U$ in order to reproduce the anomaly. The
antisymmetry under pseudo-parity conjugation can be reestablished
using that this conjugation commutes with chiral
transformations. Therefore, the following functional serves as
extended gauged Wess-Zumino-Witten (gWZW) action
\begin{equation}
\Gamma_{\text{gWZW}}[v,m] = 
\frac{1}{2}\Gamma_{\text{LR}}[v_L,v_R,\mlr]-
\frac{1}{2}\Gamma_{\text{LR}}[v_R,v_L,\mrl]\,.
\label{eq:n37}
\end{equation}
In this functional the two fields $\mlr$ and $\mrl$ are not
mixed. This is not a property of the full effective action, as it is
already clear from the form of the effective current computed in the
previous section.

In order to write this functional using our conventions, let us
consider the contribution of the (ungauged) WZW term in two dimensions
(cf. \eq{A13} setting $v$ to zero)
\begin{eqnarray}
\Gamma_{\text{WZW},d=2}[m] &=& -\frac{1}{6}\left\langle
\left(\frac{1}{\mlr}d\mlr\right)^3
-\left(\frac{1}{\mrl}d\mrl\right)^3 
 \right\rangle\,.
\label{eq:v8}
\end{eqnarray}
This can be rewritten as 
\begin{eqnarray}
\Gamma_{\text{WZW},d=2}[m] &=& \left\langle -\frac{1}{3}\sR^3
\right\rangle\,.
\label{eq:1na}
\end{eqnarray}
The meaning of the symbol $\langle\ \rangle$ was given in \eq{n2}. We
have introduced the 1-form $\sR=\frac{1}{\sm}d\sm$ and Conventions 1
and 2 apply. Note that, consistently with $\sm^{-1}\sm=1$,
$(\sm^{-1})_{LR}=\mrl^{-1}$ and $(\sm^{-1})_{RL}=\mlr^{-1}$.
More generally, in $d$ dimensions
\begin{eqnarray}
\Gamma_{\text{WZW}}[m] &=& \left\langle -\frac{1}{d+1}\sR^{d+1}
\right\rangle\,.
\label{eq:1n}
\end{eqnarray}
As usual, in $\Gamma_{\text{WZW}}[m]$ the integration takes place on a
$d+1$-dimensional disk with the $d$-dimensional space-time as
boundary. It is essential that the integrand is a closed form, so that
the result does not depend on topologically small deformations of the
$d+1$-dimensional disk. This property follows from $d\sR= -\sR^2$ and
the cyclic property. On the other hand the normalization is such that
$\Gamma_{\text{WZW}}[m]$ changes by integer multiples of $2\pi i$
under large deformations of the disk; this holds when $m$ is on the
chiral circle and the difference between $\tr(\sR^{d+1})$ on and off
the chiral circle is an exact form. (We are assuming throughout that
the fields $\mlr(x)$ and $\mrl(x)$ are nowhere singular, so any
configuration can be deformed to one on the chiral circle.)

The full gauged functional in zero, two and four dimensions, takes the form
\begin{eqnarray}
\Gamma_{\text{gWZW},d=0}[v,m] &=&
 \left\langle -\sR_c -2\sv \right\rangle \nonumber  \\ 
 &=& \left\langle -\sR \right\rangle \,, \nonumber \\
\Gamma_{\text{gWZW},d=2}[v,m] &=&
 \left\langle -\frac{1}{3}\sR_c^3
 +(\sR_c+\sL_c)\sF
+2\sv\sF - \frac{2}{3}\sv^3
 \right\rangle   \nonumber \\
&=&  \left\langle -\frac{1}{3}\sR^3 - (\sR+\sL)\sv - \sm\sv\sm^{-1}\sv
 \right\rangle \,,  \label{eq:n15}\\
\Gamma_{\text{gWZW},d=4}[v,m] 
&=&  \left\langle 
-\frac{1}{5}\sR_c^5 
+ (\sR_c^3+\sL_c^3)\sF
- 2(\sR_c+\sL_c)\sF^2
-\sR_c\sF\sm^{-1}\sF\sm -\sL_c\sF\sm\sF\sm^{-1}
\right. \nonumber \\ && \left.
 -4\sv\sF^2  +2\sv^3\sF  -\frac{2}{5}\sv^5
 \right\rangle   \nonumber \\
&=&  \left\langle 
-\frac{1}{5}\sR^5 
-(\sR^3+\sL^3)\sv
+\frac{1}{2}(\sR\sv)^2 +\frac{1}{2}(\sL\sv)^2
+\sR^2\sv\sm^{-1}\sv\sm +\sL^2\sv\sm\sv\sm^{-1}
\right. \nonumber \\ && \left.
+\sR\sm^{-1}\sv\sm d\sv +\sL\sm\sv\sm^{-1}d\sv
+(\sR+\sL)\sv^3
+\sR\sv\sm^{-1}\sv\sm\sv + \sL\sv\sm\sv\sm^{-1}\sv
\right. \nonumber \\ && \left.
+(\sR+\sL + \sm^{-1}\sv\sm + \sm\sv\sm^{-1}) \{\sv,d\sv\}
+\sm\sv\sm^{-1}\sv^3 +\sm^{-1}\sv\sm\sv^3
\right. \nonumber \\ && \left.
+\frac{1}{2}(\sm\sv\sm^{-1}\sv)^2
 \right\rangle \,.  \nonumber 
\end{eqnarray}
The functional $\Gamma_{\text{gWZW}}[v,m]$ corresponds to the LR
version of the action (as opposed to the VA version).  In these
formulas we have introduced the following 1-forms
\begin{eqnarray}
\sR &=& \sm^{-1}d\sm = -d\sm^{-1}\sm \,, \nonumber \\
\sL &=& \sm d\sm^{-1}  = -d\sm\sm^{-1} = -\sm\sR\sm^{-1}\,, \\
\sR_c &=& \sm^{-1}\sDa\sm = \sR+\sm^{-1}\sv\sm - \sv =\sm^{-1}\sm^\prime\,, \nonumber \\
\sL_c &=& \sm\sDa\sm^{-1} = -\sDa\sm\sm^{-1}  = \sL+\sm\sv\sm^{-1} - \sv = 
-\sm\sR_c\sm^{-1} =-\sm^\prime\sm^{-1} \,. \nonumber 
\end{eqnarray}
$\sR_c$ and $\sL_c$ are covariant under chiral gauge transformations.

The functional $\Gamma_{\text{gWZW}}[v,m]$ has been written in two
different forms in \eq{n15}. In the first version all terms in the
integrand are $d+1$-forms. This version shows explicitly that the
pieces which break chiral symmetry can be written as an
$\sm$-independent polynomial (in fact, this polynomial is just the
correctly normalized Chern-Simons term in $d+1$ dimensions) this
guarantees that the corresponding chiral anomaly will be also an
$\sm$-independent polynomial. Technically, such a piece looks like an
ordinary counterterm which could be removed from the effective action,
leaving a chiral invariant action. Of course, this procedure would be
incorrect, since this piece, as well as the remainder, are not
separately closed forms.  Note that the $d+1$ component of $\sv$ does
not really appear in the functional (first version) since it cancels
identically (this is easily seen in the 0-dimensional case). In the
second version, all contributions in the integrand, excepting the WZW
term, are $d$-forms. In this version the chiral symmetry is less
obvious, but it is closer to an ordinary $d$-dimensional Lagrangian.

Under the mirror transformation introduced in the previous section,
the terms which are $d$-forms are odd, whereas those written as
$d+1$-forms are even. Using $\sR\to\sL$,
$\sR_c\to\sL_c$, etc, one has, for instance,
\begin{eqnarray}
&& \langle\sR_c\sF\sm^{-1}\sF\sm\rangle \to
\langle\sm\sF\sm^{-1}\sF\sL_c\rangle =
\langle\sL_c\sm\sF\sm^{-1}\sF\rangle =
-\langle\sm\sR_c\sF\sm^{-1}\sF\rangle =
+\langle\sR_c\sF\sm^{-1}\sF\sm\rangle \,.
\end{eqnarray}

To finish this section, let us comment on the possibility of writing
the functional $\Gamma_{\text{gWZW}}[v,m]$ as a $d$-form. This has
already been done for the gauged terms. The question is whether the
WZW term $\Gamma_{\text{WZW}}[m]$, \eq{1n}, can also be written as a
$d$-form in terms of $\sm$. In Section \ref{subsec:II.B} this was done
for $\Gamma_{\text{WZW}}[U]$. The same formula does not directly apply
because $\sm$ behaves differently under the cyclic property, namely,
it is odd, whereas $U$ is even.  Indeed, if we try to use the same
method, we can see that it fails since any deformation of $\sm$ into
some $\sm(t)$ with $\sm(0)=1$ is in conflict with the condition that
the expressions must be even functions of $\sm$, for consistency. A
possibility is to go back to \eq{n37} (with $v=0$) and observe that
$\Gamma_{\text{WZW}}[\mlr]$ and $\Gamma_{\text{WZW}}[\mrl]$ are
invariant under the replacements $\mlr\to M^{-1}_{LR}\mlr$ and
$\mrl\to M^{-1}_{RL}\mrl$, where $M_{LR}$ and $M_{RL}$ are space-time
constants. Defining the new symbol $\sM$ by $(\sM)_{LR}=M_{LR}$ and
$(\sM)_{RL}=M_{RL}$, the WZW term can be written as
\begin{equation}
\Gamma_{\text{WZW}}[m] = \Gamma_{\text{WZW}}[U]\,,\quad U=\sM^{-1}\sm\,.
\end{equation}
$\sm$ and $\sM$ are odd under the cyclic property, whereas
$\sM^{-1}\sm$ is even, therefore the formulas derived in Section
\ref{subsec:II.B} apply directly.

\subsection{The chiral invariant remainder}
\label{subsec:IV.b}

Using integration by parts, the most general functionals consistent
with chiral invariance and Lorentz invariance (and at leading order in
the derivative expansion) are of the form\footnote{$W^-_c[v,m]$
vanishes in 0 dimensions. An easy calculation shows that (choosing
$\eta_0=1$) $W[v,m]=-\log(\mlr)$ and thus $\Gamma_{\text{gWZW}}[v,m]$
is the full result in this case. In addition, the most general form of
$W^-_c$ would be $\left\langle f(\sm^2) \right\rangle$, but $f$ has to
be a constant due to scale invariance, and hence it vanishes in the
pseudo-parity odd sector.}
\begin{eqnarray}
W^-_{c,d=2}[v,m] &=& \left\langle N(\sm_1,\sm_2)(\sDa\sm)^2 \right\rangle 
:= \left\langle N_{12}\sm'^2 \right\rangle 
\,, \nonumber \\
W^-_{c,d=4}[v,m] &=&  \left\langle N_{1234}\sm'^4 + 
N_{123}\sm'^2\sF \right\rangle  \,.
\label{eq:n14}
\end{eqnarray}
A comment is in order here. We have already argued above that the
current uniquely determines the effective action. Therefore, it is not
strictly necessary to deal with the most general class of chiral
invariant functionals. Any class of functionals can be used, provided
that it happens to contain the correct $W^-_c[v,m]$. The reason why
the form in \eq{n14} is sufficiently general is not entirely
straightforward, since one could imagine terms of the form $\langle
N_{1}\sF \rangle$ in two dimensions, or $\langle N_{12}\sF^2\rangle$
in four dimensions. In Section \ref{subsec:V.b} we show that those
terms are in fact redundant. On the other hand, far more general
chiral covariant terms can be devised. In \eq{n14} we have imposed
that the functional must be an analytical function of $\sm$ and
$\sD$. More general functionals exists if this condition is
lifted. However the analytical form is sufficient for the effective
action functional. Note that the analytical form comes out
automatically for the effective current, as a result of the
calculation. These more general chiral covariant functionals are
discussed in Section \ref{subsec:V.b}.

Let us discuss which restrictions exist on the functions $N$ in
$W^-_c[v,m]$.  The cyclic property implies
\begin{equation}
N_{12}= N_{2\m{1}}\,, \quad N_{1234}= N_{234\m{1}}\,. 
\label{eq:n3}
\end{equation}
(That is $N(\sm_1,\sm_2)= N(\sm_2,-\sm_1)$, etc.) This already implies
``consistency'' (i.e., the functions $N$ should be even under $\sm\to-\sm$)
as a byproduct
\begin{equation}
N_{12}= N_{\m{1}\m{2}}\,,\quad
N_{1234}= N_{\m{1}\m{2}\m{3}\m{4}}\,, \quad
N_{123}= N_{\m{1}\m{2}\m{3}}\,.
\end{equation}
Mirror symmetry requires
\begin{equation}
N_{12} = -N_{21}\,,\quad 
N_{1234} = -N_{4321}\,,\quad 
N_{123} = -N_{321}\,\,.
\end{equation}

Note the different nature of the constraints implied by the cyclic
property and mirror symmetry. Mirror symmetry is a property of our
particular functional $W^-_c[v,m]$, and it is perfectly possible to
write non-null terms violating this symmetry. On the other hand, the
cyclic symmetry is automatic; any function $N_{12}$ can be decomposed
under the group generated by $12\to 2\m{1}$, and only that component
satisfying $N_{12}= N_{2\m{1}}$ can have a non-vanishing contribution
to the functional. Thus \eq{n3} expresses our choice of working with
this relevant component only.

Dimensional counting implies that $N_{12}$ and $N_{123}$ have
dimensions of $[m^{-2}]$, and $N_{1234}$ of $[m^{-4}]$.  In addition,
the functions $N$ must be regular as two or more arguments coincide up
to a sign.

It is important to note that the functions $N$ in four dimensions are
not unambiguously determined by the functional itself, due to
integration by parts. This follows from the identity
\begin{equation}
0=  -\frac{1}{3}\left\langle \sDa\left(H_{123}\sm'^3\right)\right\rangle 
=  \left\langle -\frac{1}{3}(\Delta H)_{1234}\sm'^4 +
(\sm_1+\sm_3)H_{123}\sm'^2\sF \right\rangle \,.
\label{eq:n16}
\end{equation}
The operator $\Delta$ was defined in \eq{n4} and the identity
$\sDa^2\sm= [\sF,\sm]$ has been used. In addition the cyclic property
has been assumed on $H_{123}$. For subsequent reference we give the
cyclic property and mirror symmetry conditions on $H_{123}$
\begin{equation}
H_{123}= -H_{\m{1}\m{2}\m{3}}= -H_{23\m{1}} = -H_{321}\,.
\label{eq:n20}
\end{equation}
The identity in \eq{n16} implies that there is an ambiguity in the
definition of $N_{123}$ since it can always be augmented by
$(\sm_1+\sm_3)H_{123}$ with arbitrary $H_{123}$ subjected to the
conditions just quoted, and similarly for $N_{1234}$. (Note that
$(\Delta H)_{1234}$ does not directly have the cyclic property assumed
for $N_{1234}$, it has to be symmetrized.)

\subsection{Results}
\label{sec:4a}

First, we will need to compute the contribution of
$\Gamma_{\text{gWZW}}[v,m]$ to the effective current. In two
dimensions, a first order variation with respect to $\sv$ yields
\begin{equation}
\delta\Gamma_{\text{gWZW},d=2}[v,m]
= \langle (-\sR_c-\sL_c- 2\sv)\delta\sv\rangle, \qquad (\delta\sm=0)\,.
\end{equation}
From here, two contributions to the consistent current are identified
which correspond to the covariant contribution and the counterterm in
\eq{v10},
\begin{equation}
\sJ^{\text{WZW}}_{v,c,d=2}= -\sR_c-\sL_c\,, \qquad
 \sP_{d=2}(v)= -2\sv \,.
\label{eq:30a}
\end{equation}
Similarly, in four dimensions
\begin{eqnarray}
\sJ^{\text{WZW}}_{v,c,d=4} &=& 
-\sR_c^3-\sL_c^3
+ 2\{\sR_c+\sL_c,\sF\}
-\sm\{\sR_c,\sF\}\sm^{-1}-\sm^{-1}\{\sL_c,\sF\}\sm
\,, \nonumber \\
 \sP_{d=4}(v) &=&  4\{\sv,\sF\} -2\sv^3  \,.
\label{eq:30}
\end{eqnarray}
(The same polynomials $\sP(v)$ would be obtained with any choice of
$\Gamma_{\text{gWZW}}[v,m]$, since they are completely fixed by the
chiral anomaly.)

The contribution of $W^-_c[v,m]$ to the current is of course purely
covariant and it can be read off from
\begin{eqnarray}
\delta W^-_{c,d=2}[v,m] &=&
\left\langle -2(\sm_1+\sm_2)N_{12} \sm' \delta\sv \right\rangle
\,, \nonumber \\
\delta W^-_{c,d=4}[v,m] &=&
\Big\langle
\big(-(\Delta N)_{1234}-4(\sm_1+\sm_4)N_{1234}\big)\sm'^3\delta\sv+
\nonumber \\ &&
+\big( (\sm_1-\sm_2)N_{123}-(\sm_1+\sm_3)N_{23\m{1}}\big)\sF\sm'\delta\sv
\nonumber \\ &&
 +\big((\sm_2-\sm_3)N_{321}+(\sm_1+\sm_3)N_{21\m{3}}\big)\sm'\sF\delta\sv
\Big\rangle
 \,.
\label{eq:n12a}
\end{eqnarray}
In derivating these formulas, the cyclic property has explicitly been
assumed for the functions $N_{12}$ and $N_{1234}$. The operator
$\Delta$ was defined in \eq{n4}:
\begin{equation}
(\Delta N)_{1234} = 
 \frac{N_{134}-N_{234}}{\sm_1-\sm_2}
-\frac{N_{124}-N_{134}}{\sm_2-\sm_3}
+\frac{N_{123}-N_{124}}{\sm_3-\sm_4}\,.
\end{equation}

Collecting the different contributions to the covariant current in
eqs. (\ref{eq:30a}), (\ref{eq:30}) and (\ref{eq:n12a}), and comparing
with the definition of the functions $A$ in \eq{v7}, the following
relations are derived
\begin{eqnarray}
A_{12} &=&
- \frac{1}{\sm_1}+\frac{1}{\sm_2}-2(\sm_1+\sm_2)N_{12} 
\,, \nonumber \\
A_{123} &=&
 \frac{1}{\sm_1}+\frac{2}{\sm_2}-\frac{2}{\sm_3}-\frac{\sm_1}{\sm_2\sm_3}
+(\sm_1-\sm_2)N_{123}-(\sm_1+\sm_3)N_{23\m{1}}
\,, \nonumber \\
A_{1234} &=&
 -\frac{1}{\sm_1\sm_2\sm_3}+\frac{1}{\sm_2\sm_3\sm_4}
-(\Delta N)_{1234}-4(\sm_1+\sm_4)N_{1234}
 \,.
\label{eq:n12}
\end{eqnarray}
The terms containing $N$ are those coming from $W^-_c[v,m]$, whereas
the explicit terms are those coming from
$\Gamma_{\text{gWZW}}[v,m]$. In these relations the $N$'s are the
unknown. It is important to note that these relations have to be
augmented with the cyclicity constraints, \eq{n3}, since they have
explicitly been used in their derivation.

Let us consider the two dimensional case. For $N_{12}$ one obtains
\begin{eqnarray}
N_{12} &=&
-\frac{1}{2}\frac{1}{\sm_1+\sm_2}
\left(A_{12}+\frac{1}{\sm_1}-\frac{1}{\sm_2}\right)
\label{eq:n13} \\
&=&
-\frac{\sm_1\sm_2}{\sm_1^2-\sm_2^2}\left(\frac{\log(\sm_1^2/\sm_2^2)}
{\sm_1^2-\sm_2^2}-\frac{1}{2}\left(
\frac{1}{\sm_1^2}+\frac{1}{\sm_2^2}\right)\right)
\,.
\label{eq:n13a}
\end{eqnarray}
It is worth noticing that the correct cyclic property for $N_{12}$,
namely, $N_{12}=N_{2\m{1}}$, is verified, but this is not an automatic
consequence of \eq{n13}. This poses a severe restriction on the a
priori admissible functions $A_{12}$, if they should derive from an
effective action with the correct Lorentz and chiral symmetries. In
addition the function $N_{12}$ is finite, i.e. regular at
$\sm_1^2=\sm_2^2$. Again this property does not follow automatically
from finiteness of $A_{12}$. On the other hand, scale and mirror
symmetries are automatic in $N_{12}$ from the corresponding symmetries
in $A_{12}$.

Another comment is that in the first \eq{n12} (similar remarks apply
to the four-dimensional formulas) the WZW contribution to $A_{12}$
cannot be reabsorbed into the contribution coming from $W^-_c[v,m]$ by
means of a suitable redefinition of the function $N_{12}$. If this
were the case, we would have that the covariant current, $A_{12}$, is
also consistent (it would derive from a certain
$W^-_c[v,m]$). Technically the reason is that such a function $N_{12}$
would violate the cyclic property constraint. (In addition it would
not be finite at $\sm_2=-\sm_1$.) Thus in this formalism the
non-integrability of the covariant current, which necessarily implies
the existence of a chiral anomaly, translates into a breakdown of the
cyclic property.

In summary, the function $N_{12}$ in \eq{n13a} inserted in
$W^-_c[v,m]$ in \eq{n14} plus the extended gauged WZW term in \eq{n15}
provides the full functional for the leading order of the
pseudo-parity odd component of the effective action in two dimensions.

Let us now turn to the four dimensional case. Unfortunately, this case
is more involved, mainly because of the presence of ambiguities in the
functions $N$ introduced by integration by parts. These ambiguities do
not affect the functional $W^-_c[v,m]$ itself.

Mirror symmetry of $N_{123}$ automatically implies
$\bar{A}_{123}=\bar{A}_{132}$, which can thus be understood as a
consequence of mirror plus Lorentz symmetries (chiral symmetry is not
required). The full permutation symmetry of $\bar{A}_{123}$ and
$\bar{A}_{1234}$ would not follow if $N_{123}=N_{1234}=0$ and so it
cannot be understood in this way.

Clearly $N_{1234}$ is uniquely determined by the formulas once
$A_{1234}$ and $N_{123}$ are known. However, $N_{123}$ is
not unambiguously determined from $A_{123}$. In turn this puts a
restriction on the possible $A_{123}$, namely,
\begin{equation}
(\sm_2-\sm_3)A_{123}+ (\sm_1+\sm_3)A_{23\m{1}}+
(\sm_2-\sm_1)A_{3\m{1}\m{2}} = 12 \,,
\end{equation}
which is verified by the true function $A_{123}$. This ambiguity was
noted above, \eq{n16}. It is verified that the modification introduced
by $H_{123}$ exactly cancels in the right-hand side of \eq{n12}. This
serves as a check of these formulas. This means that the functions
$N_{123}$ and $N_{1234}$ are ambiguous but not the functional
$W^-_c[v,m]$ itself. This is consistent with the fact that the current
completely fixes the effective action functional if the correct chiral
transformation is assumed.

A particular solution for the functional $N_{123}$ is given by
\begin{equation}
N^0_{123} = 
\frac{1}{3}\left(
\frac{A_{123}}{\sm_1-\sm_2} + \frac{A_{3\m{1}\m{2}}}{\sm_2-\sm_3}
\right)
-\frac{1}{(\sm_1-\sm_2)(\sm_2-\sm_3)}
\left(\frac{\sm_1}{\sm_2}+\frac{\sm_2}{\sm_1}
-\frac{\sm_2}{\sm_3}-\frac{\sm_3}{\sm_2}\right) \,.
\label{eq:77}
\end{equation}
This is easily verified by substitution. The associated function
$N^0_{1234}$ is immediately obtained from \eq{n12}. Besides the
trivial mirror symmetry and scale invariances, it is verified that
$N^0_{1234}$ possesses the correct cyclic symmetry. Again this is a
highly non-trivial check of the functions $A_{123}$ and $A_{1234}$.
However, the functions $N^0_{123}$ and $N^0_{1234}$ are not directly
acceptable since they fail to be finite in the coincidence limit,
namely, when $\sm_1=\sm_2$ or $\sm_2=\sm_3$. This implies that another
solution has to be chosen by taking an appropriate function
$H_{123}$. (Note that the previous checks are preserved by this
operation.)

To find an acceptable solution it is convenient to work with a reduced
version of the function $N_{123}$, namely
\begin{equation}
\widehat{N}_{123}=(\sm_1-\sm_2)(\sm_2-\sm_3)N_{123}\,.
\label{eq:n22}
\end{equation}
Consistency and mirror symmetry of $N_{123}$ translate into
\begin{equation}
\widehat{N}_{123} = \widehat{N}_{\m{1}\m{2}\m{3}}=-\widehat{N}_{321}\,.
\end{equation}
On the other hand, the condition of finiteness of $N_{123}$ at
$\sm_1=\sm_2$ corresponds to
\begin{equation}
\widehat{N}_{113}= 0 \,.
\label{eq:n34}
\end{equation}
Due to mirror symmetry this immediately implies $\widehat{N}_{122}=0$ and thus
finiteness of $N_{123}$ at $\sm_2=\sm_3$ too. This finiteness
condition is violated by $\widehat{N}^0_{123}$.

Likewise, for the function $H_{123}$ controlling the ambiguity we
define its reduced version as
\begin{equation}
\widehat{H}_{123}= (\sm_1-\sm_2)(\sm_2-\sm_3)(\sm_1+\sm_3)H_{123}\,.
\end{equation}
Consistency, cyclic property and mirror symmetry of $H_{123}$ in
\eq{n20} translate into
\begin{equation}
\widehat{H}_{123}= \widehat{H}_{\m{1}\m{2}\m{3}}= 
\widehat{H}_{23\m{1}} = -\widehat{H}_{321}\,.
\label{eq:n45a}
\end{equation}
(This is equivalent to say that the function $\widehat{H}_{1\m{2}3}$
is completely antisymmetric under permutation of its arguments and
even under $\sm\to-\sm$.)

In terms of the reduced functions, the ambiguity corresponds to the
fact that $\widehat{N}_{123}$ and $\widehat{N}_{123}-\widehat{H}_{123}$ produce the same current
$A_{123}$. In view of this, our strategy is to find an $\widehat{H}_{123}$ such
that $\widehat{H}_{113}=\widehat{N}^0_{113}$, so that
\begin{equation}
\widehat{N}_{123} =  \widehat{N}^0_{123} - \widehat{H}_{123}
\label{eq:n23}
\end{equation}
fulfills the finiteness condition, \eq{n34}. This can be done as
follows.  Although the function $\widehat{N}^0_{123}$ does not vanish
at $\sm_1=\sm_2$, it is finite and satisfies
\begin{equation}
\widehat{N}^0_{123}=\widehat{N}^0_{\m{1}\m{2}\m{3}}\,,\quad
\widehat{N}^0_{111}=0\,,\quad
\widehat{N}^0_{113}= -\widehat{N}^0_{11\m{3}}\,.
\label{eq:n21}
\end{equation}
The first relation is consistency. The second one comes from mirror
symmetry and the last one follows from finiteness of $A_{123}$ (this
is more simply verified from \eq{n12}). These relations imply that the
function
\begin{equation}
\widehat{H}_{123}= \frac{1}{2}\left( -\frac{\sm_3}{\sm_1}\widehat{N}^0_{112}
+\frac{\sm_1}{\sm_2}\widehat{N}^0_{223} -\frac{\sm_2}{\sm_3}\widehat{N}^0_{331}
-\frac{\sm_3}{\sm_2}\widehat{N}^0_{221} +\frac{\sm_1}{\sm_3}\widehat{N}^0_{332}
+\frac{\sm_2}{\sm_1}\widehat{N}^0_{113} \right)\,
\label{eq:84}
\end{equation}
satisfies the requirements in \eq{n45a}. In addition
\begin{equation}
\widehat{H}_{113}= \frac{1}{2}\left( -\frac{\sm_3}{\sm_1}\widehat{N}^0_{111}
+\widehat{N}^0_{113} -\frac{\sm_1}{\sm_3}\widehat{N}^0_{331}
-\frac{\sm_3}{\sm_1}\widehat{N}^0_{111} +\frac{\sm_1}{\sm_3}\widehat{N}^0_{331}
+\widehat{N}^0_{113} \right) 
= \widehat{N}^0_{113}\,,
\end{equation}
therefore the function $\widehat{N}_{123}$ defined as $\widehat{N}^0_{123} -
\widehat{H}_{123}$ automatically satisfies $\widehat{N}_{113}=0$ and thus it yields a
finite $N_{123}$. Finiteness of the corresponding $N_{1234}$ also
follows automatically: because $N_{123}$ and $A_{1234}$ are finite,
\eq{n12} implies the $N_{1234}$ is also finite except perhaps at
$\sm_1=-\sm_4$, however, this follows from the cyclic property,
$N_{1234}=N_{\m{4}123}$ and finiteness of $N_{1123}$.

Let us summarize the result. The acceptable $N_{123}$ is obtained as
follows: from $A_{123}$ (\eq{69b}), one obtains $N^0_{123}$ (\eq{77}),
then $\widehat{N}^0_{123}$ (\eq{n22}) and $\widehat{H}_{123}$ (\eq{84}). This gives
$\widehat{N}_{123}$ (\eq{n23}) and $N_{123}$ (\eq{n22}). Finally, $N_{1234}$
follows from \eq{n12}. The explicit resulting functions are displayed
in Appendix \ref{app:B}.

It can be noted that in addition to the redefinition from $N^0_{123}$
to $N_{123}$ to achieve finiteness, further redefinitions, by suitable
finite functions $H_{123}$, can be made to simplify the final form of
$N_{123}$ and $N_{1234}$. In practice, we have not been able to
achieve a greater simplification.  Certainly the functions $N_{123}$
and $N_{1234}$ cannot be much simpler than the functions $A_{123}$ and
$A_{1234}$, which are free from ambiguities, thus no simple form is to
be expected for the functions $N$.

As we have seen in subsection \ref{subsec:II.B}, the WZW term has a
simple form when written as a $d+1$-dimensional integral but looks
complicated in terms of $d$-forms. One can wonder whether this is also
the case for $W^-_c[v,m]$. Applying the operator $\sDa$ to its
integrand, $W^-_c[v,m]$ can be written as a $d+1$-form, however no
simplification occurs. Again, a large simplification would have been in
contradiction with the unambiguous form of the effective current.

\section{Further comments and results}
\label{sec:5}

\subsection{The chiral circle constraint}
\label{subsec:V.a}

The previous calculations are completely general regarding the chiral
group and the external field configurations since no assumption has
been made on the algebraic properties in flavor space.  Let us now
discuss the form of the functional on the chiral circle.  A field
configuration $(v,m)$ is on the chiral circle when $\mlr(x)=MU(x)$ and
$\mrl(x)=MU^{-1}(x)$ where $M$ is constant c-number. By unitarity $U$
must be a unitary matrix, but in practice we will only use that $U$ is
nowhere singular. Due to dimensional counting $M$ cannot appear in
$W^-[v,m]$ (since we are considering the leading term only and all
dimensions are already accounted for by the derivatives and the gauge
fields) thus we can take $M=1$ and express the chiral circle
constraint as $\mlr\mrl=1$ or equivalently as $\sm^2= 1$.

As is well-known, on the chiral circle the leading term of $W^-[v,m]$
is saturated by the gauged WZW action $\Gamma_{\text{LR}}[v,U]$. This
comes about because it is possible to chirally rotate the
configuration by $U$ to bring it to the form $\mlr =\mrl=1$ and so
$W^-[v,m]$ is given by $\Gamma_{\text{LR}}[v,U]$ plus
$W^-_{\text{VA}}[v,m=1]$ (see Appendix \ref{app:A}). Thus the
statement is equivalent to saying that the leading term of
$W^-_{\text{VA}}[v,m]$ vanishes when $\mlr =\mrl=1$. This follows
because the possible vector gauge invariant terms constructed out of
$v_L$ and $v_R$ of dimension $d$ vanish identically. (Vector gauge
invariance is the remaining chiral invariance compatible with the
condition $\mlr =\mrl=1$, and it must be preserved since all the
anomaly is saturated by the gauged WZW term).

The fact that, on the chiral circle at leading order in the derivative
expansion, $W^-[v,m]= \Gamma_{\text{LR}}[v,U]$, is of course contained
in our general formulas. Actually, a stronger statement can be
deduced, namely, the leading order of $W^-[v,m]$ is saturated by
$\Gamma_{\text{LR}}[v,U]$ whenever 
\begin{equation}
\mlr(x)=M(x)U(x)\,, \qquad \mrl(x)=M(x)U^{-1}(x)
\label{eq:n35}
\end{equation}
where $M(x)$ is a c-number but not a necessarily constant. (Note that
this class of configurations is closed under chiral transformations.)
To show this, let us define the symbol $\sM$ by
$(\sM)_{RR}=(\sM)_{LL}=M$ and $\sU$ by $(\sU)_{LR}=U$ and
$(\sU)_{RL}=U^{-1}$. (Note that $\sU^2=1$ and so $\sU=\sU^{-1}$.) This
allows to use the Convention 2: $\sm= \sM\sU$ and
$\sm'=d\sM\sU+\sM\sU'$.

Consider first the extended WZW term.  Clearly, in
$\Gamma_{\text{gWZW}}[v,m]$ all dependence on $\sM$ without
derivatives cancels, by simple dimensional counting. Likewise, all
terms with two or more $d\sM$ also cancel trivially since $d\sM$ is a
c-number and $(d\sM)^2=0$. Finally, the terms linear in $d\sM$ can be
shown to cancel too, by using (recall that $\sU= \sU^{-1}$)
\begin{equation}
\sR_c=\sR^U_c+\sR_M\,, \quad \sL_c=\sR^U_c-\sR_M\,,\quad
\sR_c^U=\sU\sU'\,,\quad \sR_M=\sM^{-1}d\sM \,.
\end{equation}
For instance $\sR_c^5= (\sR_c^U)^5+(\sR_c^U)^4\sR_M$ and $\sL_c^5=
(\sR_c^U)^5-(\sR_c^U)^4\sR_M$, thus $\sR_c^5+\sL_c^5=2(\sR^U_c)^5$.
Therefore, when $\sm=\sM\sU$
\begin{equation}
\Gamma_{\text{gWZW}}[v,m]=\Gamma_{\text{LR}}[v,U] \,,
\end{equation}
this is the same as the chiral circle result.

Let us now show that $W^-_c[v,m]$ vanishes identically. In fact this
holds not only for the true functional $W^-_c[v,m]$ but also for any
other finite functional with the correct symmetries. Therefore only
general properties of the functions $N$ are needed in the proof.

Consider first the term with $N_{12}$. Due to scale invariance
\begin{equation}
N(\sm_1,\sm_2)= \frac{1}{\sM^2}N(\sU_1,\sU_2)\,.
\end{equation}
Because $\sU^2=1$, the function $N(\sU_1,\sU_2)$ is completely
equivalent
to one where each of the $\sU_{1,2}$ is raised to the first power, at most: 
\begin{equation}
N(\sU_1,\sU_2)= a+b\sU_1+c\sU_2+d\sU_1\sU_2\,,
\end{equation}
where $a,b,c,d$ are some constants (these constants exist since
$N_{12}$ is finite in the coincidence limit). However, consistency
requires $b=c=0$ ($N_{12}$ is an even function of $\sm$). Further,
mirror symmetry requires $a=d=0$ too, and $W^-_c[v,m]$ vanishes
identically in two dimensions for configurations of the form
$\sm=\sM\sU$.

In four dimensions, scale invariance, consistency and the cyclic
property imply
\begin{equation}
N(\sm_1,\sm_2,\sm_3,\sm_4)= \frac{1}{\sM^4}
(a+b(\sU_1\sU_2+\sU_2\sU_3+\sU_3\sU_4-\sU_4\sU_1) )
\end{equation}
for some constants $a$ and $b$. However mirror symmetry requires
$a=b=0$. Thus there is no contribution from $\langle N_{1234}\sm'^4\rangle$.

The term $\langle N_{123}\sm'^2\sF\rangle$ is slightly more
complicated. In this case scale invariance, consistency and mirror
symmetry imply
\begin{equation}
N(\sm_1,\sm_2,\sm_3)= \frac{a}{\sM^2} (\sU_1-\sU_3)\sU_2 \,.
\end{equation}
The constant $a$ needs not vanish (in fact, $a=-\frac{1}{2}$ for the
true functional). Nevertheless, a straightforward calculation using
$\sm'=d\sM\sU+\sM\sU'$, $\sU\sU'= -\sU'\sU$ and that $\sM$ is a
c-number, shows that this contribution vanishes as well.

It is also worth point out that on the strict chiral circle, i.e. $M$
constant, $W^-_c[v,m]$ can be shown to vanish without assuming mirror
symmetry.

Another remark is that the previous statements also hold for any
Abelian theory, i.e. when all matrices are c-numbers in flavor
space. This is because in this case $\mlr$ and $\mrl$ can certainly be
written as in \eq{n35} with $M(x)$ a c-number, so the previous results
apply.

After all these null results, one could wonder whether the chiral
invariant remainder is not actually identically zero, although this is
not obvious due to the notation. We have explicitly verified that this
is not the case using a two-flavor model in two dimensions without
accidental symmetries.

\subsection{The effective density}
\label{subsec:V.e}
In this subsection we give explicit formulas for the effective density
$\sJ^-_m$ introduced in \eq{curr} as the variation of the effective
action with respect to $\sm$. The general form of the densities is
\begin{eqnarray}
\sJ^-_{m,d=2} &=& B_{123}\sm'^2+B_{12}\sF\,, \nonumber \\
\sJ^-_{m,d=4} &=& B_{12345}\sm'^4 +
B_{1234}\sm'\sF\sm'+ B'_{1234}\sm'^2\sF+ B''_{1234}\sF\sm'^2
+ B'_{123}\sF^2 \,.
\end{eqnarray}
For consistency, the functions $B$ are all odd under $\sm\to -\sm$. In
addition mirror symmetry implies
\begin{eqnarray} 
&& 
B_{12}=B_{21}\,,\quad
B_{123}=B_{321}\,, \nonumber \\ &&
B'_{123}=B'_{321}\,,\quad
B_{1234}=B_{4321}\,,\quad
B'_{1234}=B''_{4321}\,,\quad
B_{12345}=B_{54321}\,.
\end{eqnarray}

The effective density can be computed from scratch, by the same method
used in Section \ref{subsec:III.b} for the effective current.  Within
our approach, the direct calculation of the density is harder than for
the current because they are of higher order ( $\sJ^-_m$ is a $d$-form
whereas $\sJ^-_v$ is a $d-1$-form). A better procedure is to obtain
the effective density as the variation of the effective action, which
has already been computed. As noted the consistent effective density
is also covariant. 

An explicit variation of $\Gamma_{\text{gWZW}}[v,m]$ and $W^-_c[v,m]$
in two dimensions yields
\begin{eqnarray}
B_{12} &=&
 - 2 (\sm_1-\sm_2) N_{12} +\frac{1}{\sm_1}  +\frac{1}{\sm_2} \,,
\nonumber \\
B_{123} &=&
   2 \left(
    \frac{N_{13}-N_{23}}{\sm_1-\sm_2}
  - \frac{N_{12}-N_{13}}{\sm_2-\sm_3}
  + \frac{N_{23}-N_{12}}{\sm_3+\sm_1}
    \right)
   - \frac{1}{\sm_1 \sm_2 \sm_3} \,.
\label{eq:density2}
\end{eqnarray}
The terms with $N$ are those coming from $W^-_c[v,m]$; the other come
from $\Gamma_{\text{gWZW}}[v,m]$.

In four dimensions
\begin{eqnarray}
B'_{123} &=&
   -(\sm_1-\sm_2)N_{\m{3}12}-(\sm_2-\sm_3)N_{23\m{1}}
   - \frac{2}{\sm_3}-\frac{1}{\sm_2}-\frac{2}{\sm_1}
-\frac{\sm_2}{\sm_1\sm_3} \,,
\nonumber \\
B_{1234} &=&
     4 (\sm_2-\sm_3) N_{1234}
   - \frac{N_{\m{3}12}-N_{\m{4}12}}{\sm_3-\sm_4}
   + \frac{N_{34\m{1}}-N_{34\m{2}}}{\sm_1-\sm_2}
   + \frac{N_{34\m{2}}-N_{\m{3}12}}{\sm_4+\sm_1}
\nonumber \\ && 
   + \frac{1}{\sm_1 \sm_3 \sm_4}+\frac{1}{\sm_1 \sm_2 \sm_4} \,,
\nonumber \\
B'_{1234} &=&
   - 4 (\sm_3-\sm_4) N_{1234}
   + \frac{N_{\m{4}13}-N_{\m{4}23}}{\sm_1-\sm_2}
   - \frac{N_{\m{4}12}-N_{\m{4}13}}{\sm_2-\sm_3}
   + \frac{N_{\m{4}23}-N_{123}}{\sm_4+\sm_1}
\nonumber \\ && 
   + \frac{1}{\sm_1 \sm_2 \sm_4}+\frac{1}{\sm_1 \sm_2 \sm_3} \,,
\nonumber \\
B_{12345} &=&
   4 \Bigg(
    \frac{N_{1345}-N_{2345}}{\sm_1-\sm_2}
  - \frac{N_{1245}-N_{1345}}{\sm_2-\sm_3}
  + \frac{N_{1235}-N_{1245}}{\sm_3-\sm_4}
  - \frac{N_{1234}-N_{1235}}{\sm_4-\sm_5}
\nonumber \\ && \quad
  + \frac{N_{2345}-N_{1234}}{\sm_5+\sm_1}
    \Bigg)
   - \frac{1}{\sm_1 \sm_2 \sm_3 \sm_4 \sm_5}\,.
\label{eq:density4}
\end{eqnarray}

There is an alternative way to obtain the density which is simpler and
also serves as a check, namely by using the anomaly equation. This is
\eq{curr} when the variations are associated to an infinitesimal
chiral rotation \eq{7b}
\begin{equation}
\sDa\sJ_v^- + \{\sJ_m^-,\sm\} =  \Ar \,,
\label{eq:112}
\end{equation}
where $\Ar$ is the consistent chiral anomaly (defined so that $\delta
W^-_{\text{LR}}[v,m]=\langle \Ar\alpha\rangle$ is the left-hand side
of \eq{27}). Note that $\sJ_v^-$ is the consistent current. The
contribution of the counterterm current $\sP(v)$ cancels the chiral
symmetry breaking terms from the anomaly. This yields the following
formulas for the density
\begin{eqnarray}
B_{12}&=& \frac{1}{\sm_1+\sm_2}(4+(\sm_1-\sm_2)A_{12}) \,, \nonumber
\\ B_{123}&=& -\frac{1}{\sm_1+\sm_3}(\Delta A)_{123} \,, \nonumber \\
B^\prime_{123}&=& -\frac{1}{\sm_1+\sm_3} (12 -(\sm_2-\sm_3)A_{123}
+(\sm_1-\sm_2)A_{321} ) \,, \nonumber \\ B_{1234}&=&
-\frac{1}{\sm_1+\sm_4} ( \frac{A_{134}-A_{234}}{\sm_1-\sm_2}
+\frac{A_{321}-A_{421}}{\sm_3-\sm_4} +(\sm_2-\sm_3)A_{1234} ) \,,
\nonumber \\ B^\prime_{1234}&=& -\frac{1}{\sm_1+\sm_4} (
-\frac{A_{431}-A_{432}}{\sm_1-\sm_2}
+\frac{A_{421}-A_{431}}{\sm_2-\sm_3} -(\sm_3-\sm_4)A_{1234} ) \,,
\nonumber \\ B_{12345}&=& -\frac{1}{\sm_1+\sm_5}(\Delta A)_{12345} \,.
\label{eq:densityA}
\end{eqnarray}
It can be verified that these expressions coincide with those in
eqs. ({\ref{eq:density2}) and ({\ref{eq:density4}) and the possible
ambiguities introduced by the functions $N$ are explicitly removed.

As in the case of the effective current we can define a set of
associated functions as follows
\begin{eqnarray}
B_{12} &=& \bar{B}_{12}\,,\quad
B_{123}= \bar{B}_{1\m{2}3}\,,
\nonumber \\
B'_{123} &=& \bar{B}'_{123}\,,\quad
B_{1234}= \bar{B}_{1\m{2}\m{3}4}\,,\quad
B'_{1234}= \bar{B}'_{1\m{2}34}\,,\quad
B_{12345}= \bar{B}_{1\m{2}3\m{4}5} \,.
\end{eqnarray}
(The rule is to flip the signs of the arguments at the right of each
operator that is an odd-order differential form, in practice $\sm'$.)
Once again the functions $\bar{B}$ so defined turn out to have the
property of being completely symmetric under permutation of their
arguments, a property already noted for the functions $\bar{A}$ of the
effective current. In two dimensions this property follows solely from
the general symmetries of the function $N_{12}$, however, in four
dimensions this is not the case. Consistency and mirror symmetry
follows automatically in all cases from the corresponding properties
of $N_{123}$ and $N_{1234}$. Invariance of $ \bar{B}'_{123}$,
$\bar{B}_{1234}$, and $\bar{B}'_{1234}$ under general permutations
(other than mirror permutations) does not follow from general
symmetries of $N_{123}$ and $N_{1234}$ as is already obvious by
setting these two functions to zero in \eq{density4}. For
$\bar{B}_{12345}$, it can be shown that invariance under cyclic
permutations follows from general symmetries of $N_{123}$ and
$N_{1234}$ but invariance under more general permutations does not.
(Also the complete symmetry of the functions $\bar{A}$ combined with
the formulas in \eq{densityA} does not guarantee symmetry of the
functions $\bar{B}$.) Therefore, the complete symmetry of the
functions $\bar{B}$ in four dimensions is a specific property of the
true effective action functional. Since this property is so general
(it holds for effective currents and effective densities and in all
dimensions examined) it is likely that it follows from the very
definition of these currents rather than being an accidental symmetry.

\subsection{Vector-like reduction}
\label{subsec:V.f}

Our conventions for the  vector-axial (VA) notation are as follows
\begin{equation}
\Dirac= \thru{\D}_V+\thru{A}\gamma_5+ S+{\gamma_5}P\,,
\end{equation}
where $\D^V_\mu=\partial_\mu+V_\mu$ is the vector covariant derivative
and
\begin{equation}
v_{R,L}= V\pm A\,,\quad \mlr = S+P\,,\quad \mrl = S-P\,.
\end{equation}
Strictly speaking a purely vector-like case would mean $v_R=v_L$ and
$\mlr=\mrl$, or $P=A=0$. For such configurations there is no
pseudo-parity odd component of the effective action. Thus, presently,
by vector-like case we will refer to the case of vanishing
pseudo-scalar field, $P=0$, but not necessarily vanishing axial field
$A$. Of course, in this case it is preferable to work with the VA
version of the effective action, which is related to the LR version by
subtracting an appropriated $m$-independent counter-term (see Appendix
\ref{app:A})
\begin{equation}
W^-_{\text{VA}}[v,m] = W^-_{\text{LR}}[v,m] - P_{\text{ct}}[v]\,.
\label{eq:repe}
\end{equation}
The counter-term is such that $W^-_{\text{VA}}[v,m]$ is vector gauge
invariant and the anomaly affects only axial transformations. In this
subsection we will denote $W^-[v,m]$ by $W^-_{\text{LR}}[v,m]$ to
emphasize that it is the LR version of the effective action.

When $P=0$, the most general form of the VA effective action (at
leading order and in the pseudo-parity odd sector) is
\begin{eqnarray}
W^-_{\text{VA},d=2}[v,m] &=&
\Big\langle M_{12}S'A \Big\rangle\,,
\label{eq:n134}
\\
W^-_{\text{VA},d=4}[v,m] &=&
\Big\langle
M_{123}F_VS'A + M'_{123}S'F_VA + M_{1234}S'^3A + 
M'_{1234}S'A^3 + M''_{123}A^2F_A \Big\rangle\,,
\nonumber
\end{eqnarray}
where $S'=[D_V,S]$, $F_V=D_V^2$ and $F_A=\{D_V,A\}$ and the various
$M$'s are functions of $S$, i.e, $M_{12}=M(S_1,S_2)$, etc.  Note that
the symbol $S$, unlike $\sm$, is even under cyclic permutations, thus
in particular there is no consistency restrictions on the various
functions $M$. Also there are no cyclicity restrictions. On the other
hand, mirror symmetry is guaranteed provided that
\begin{equation}
M_{12}=-M_{21}\,,\quad
M_{123}=-M'_{321}\,,\quad
M_{1234}=-M_{4321}\,,\quad
M'_{1234}=-M'_{2143}\,,\quad
M''_{123}=-M''_{321}\,.
\end{equation}

Not all these functions are unambiguously determined by the functional
itself due to the following identity (the prime denotes derivative
with respect to $D_V$)
\begin{equation}
0= \frac{1}{3}\Big\langle  (G_{123}A^3)' \Big\rangle
=\Big\langle  G_{123}A^2F_A +\frac{G_{134}-G_{234}}{S_1-S_2}S'A^3 \Big\rangle
\,.
\end{equation}
Here $G_{123}$ is any function subjected to the cyclic property
restriction $G_{123}=G_{231}$. If in addition mirror symmetry is
imposed, $G_{123}$ must be a completely antisymmetric function under
permutation of its arguments. This identity introduces an integration
by parts ambiguity in $M'_{1234}$ and $M''_{123}$.

Since $W^-_{\text{LR}}[v,m]$ has been computed previously, \eq{repe}
can be used to obtain $W^-_{\text{VA}}[v,m]$. The contribution from
$\Gamma_{\text{gWZW}}[v,m]$ when $P=0$ is easily obtained from
\eq{n37}. This contribution combined with that coming from the
counterterm yields
\begin{eqnarray}
W^-_{\text{VA,WZW},d=2}[v,m] &=&\Big\langle -[S^{-1},S']A\Big\rangle 
\nonumber \\
W^-_{\text{VA,WZW},d=4}[v,m] &=&\Big\langle
2F_V[S^{-1},S']A +S^{-1}F_VS'A-SF_VS^{-1}S'S^{-1}A
\nonumber \\ && 
+2[S^{-1},S']F_VA -S'F_VS^{-1}A+S^{-1}S'S^{-1}F_VSA
\nonumber \\ && 
-(S^{-1}S')^3A +(S'S^{-1})^3A
\nonumber \\ && 
+[S^{-1},S']A^3+S'S^{-1}ASAS^{-1}A-S^{-1}S'AS^{-1}ASA
\nonumber \\ && 
-SAS^{-1}AF_A -S^{-1}ASAF_A +ASAS^{-1}F_A +AS^{-1}ASF_A
\Big\rangle \,.
\end{eqnarray}
By construction all terms breaking vector gauge invariance have
canceled.

The contribution coming from $W^-_c[v,m]$ is also easily computed. To
illustrate the method we will work out explicitly the two-dimensional
case. From consistency, the most general form of $N_{12}$ is
\begin{equation}
N_{12}= n'_{12}+\sm_1\sm_2 n_{12}\,,
\label{eq:n138}
\end{equation}
where $n_{12}$ and $n'_{12}$ are functions of $\sm_1^2$ and $\sm_2^2$.
(Mirror symmetry further requires $n'_{12}$ to vanish but this will
not be enforced here.) Thus (expanding Convention 2 but keeping
Convention 1) yields
\begin{eqnarray}
\Big\langle N_{12}\sm'^2\Big\rangle &=&
\Big\langle (n'_{12}+\sm_1\sm_2 n_{12})\sm'^2 \Big\rangle \\
&=& \Big\langle n'_{12}\mrl'\mlr'+n_{12}(\sm\sm')_R^2 \Big\rangle \,.
\end{eqnarray}
Now we can take $P=0$, i.e., $\mlr=\mrl=S$ and use the formulas
\begin{equation}
\mrl'= S'+\{S,A\}\,,\quad \mlr'= S'-\{S,A\}\,.
\end{equation}
(Note that the prime refers to $\D_{R,L}$ in the left-hand side and to
$D_V$ in the right-hand side.) In addition, all arguments $\sm$ become
$S$. This gives
\begin{eqnarray}
W^-_{\text{VA},c,d=2}[v,m]
&=& \Big\langle n'_{12}(S'+\{S,A\})(S'-\{S,A\})
+n_{12}S_1S_2(S'-\{S,A\})(S'-\{S,A\}) \Big\rangle
\nonumber \\
&=& \Big\langle -n'_{12}S'\{S,A\} + n'_{12}\{S,A\}S'
-n_{12}S_1S_2S'\{S,A\} - n_{12}S_1S_2\{S,A\}S'  \Big\rangle
\nonumber \\
&=& \Big\langle -N_{12}S'\{S,A\} + N_{1\m{2}}\{S,A\}S' \Big\rangle
\nonumber \\
&=& \Big\langle -2N_{12}S'\{S,A\} \Big\rangle \,.
\end{eqnarray}
The second equality follows from Convention 1 which keeps only
pseudo-parity odd terms. The last equality follows from the cyclic
property of the trace. A useful observation is that, although the
detailed expansion in \eq{n138} is required in intermediate steps, the
full function $N_{12}$ can be reconstructed, as in the third line
above, by allowing appropriate changes in the signs of its arguments
(e.g. $N_{1\m{2}}$).\footnote{The empirical rule is to flip the signs
of the arguments at the right of each $A$ or $F_A$. In addition there
is a global minus sign for each $A$ or $F_A$ occupying an even
position.}

In four dimensions, using
\begin{equation}
F_R= F_V+A^2+F_A\,,\quad F_L= F_V+A^2-F_A\,,
\end{equation}
the result is
\begin{eqnarray}
W^-_{\text{VA},c,d=4}[v,m]
&=& \Big\langle N_{\m{1}23}\{S,A\}S'(F_V+A^2)
-N_{12\m{3}}S'\{S,A\}(F_V+A^2)
+N_{123}S'^2F_A
\nonumber \\ &&
-N_{1\m{2}3}\{S,A\}^2F_A
-4N_{1234}S'^3\{S,A\}
+4N_{12\m{3}4}S'\{S,A\}^3
\Big\rangle \,.
\end{eqnarray}
In this formula, the term $S'^2F_A$ has to be integrated by parts in
order to conform to the standard form chosen in \eq{n134}.

The combination of the previous results from
$\Gamma_{\text{gWZW}}[v,m]$, $W^-_c[v,m]$ and $P_{\text{ct}}[v]$
gives
\begin{eqnarray}
M_{12} &=& A_{12}\,,\nonumber \\
M_{123} &=& A_{123}\,,\nonumber \\
M_{1234} &=& A_{1234}\,,\nonumber \\
M'_{1234} &=& 
\frac{1}{S_1}-\frac{1}{S_2}+\frac{S_3}{S_2S_4}-\frac{S_4}{S_1S_3}
-(S_2+S_3)N_{12\m{3}} -(S_1+S_4)N_{\m{4}12} \nonumber \\
&& +4(S_1+S_4)(S_2+S_3)(S_3+S_4)N_{12\m{3}4} \,,\nonumber \\
M''_{123} &=& 
-\frac{S_1}{S_2}-\frac{S_2}{S_1}+\frac{S_2}{S_3}+\frac{S_3}{S_2}
-(S_1+S_2)(S_2+S_3)N_{1\m{2}3} \,.
\label{eq:145}
\end{eqnarray}
In these formulas the functions $A_{12}$, etc, are those of the
effective current in \eq{v7}.

The ambiguity in the functions $N_{123}$ and $N_{1234}$ translates
into an ambiguity in $M'_{1234}$ and $M''_{123}$ of the form $G_{123}=
-\widehat{H}_{1\m{2}3}$. On the other hand, exploiting the ambiguity
in these functions allows to write explicit expressions in terms of
the $A$'s as follows
\begin{eqnarray}
M'_{1234} &=& 
-\frac{1}{3}\frac{(S_1+S_4)A_{143}-(S_2+S_4)A_{243}}{S_1-S_2}
-\frac{1}{3}\frac{(S_1+S_3)A_{\m{1}43}-(S_2+S_3)A_{\m{2}43}}{S_1-S_2}
\nonumber \\
&&-(S_1+S_4)(S_2+S_3)A_{12\m{3}\m{4}}\,,
\nonumber \\ 
M''_{123} &=& 
\frac{1}{3}(S_1+S_2)A_{3\m{1}2} 
-\frac{1}{3}(S_2+S_3)A_{1\m{3}2}\,.
\label{eq:144}
\end{eqnarray}
Note that these functions differ from those in \eq{145}, although, of
course, they produce the same functional.

An alternative way to obtain the functions $M$ is based in reproducing
the correct axial current. This method yields \eq{144} more
directly. The procedure is straightforward, so we do not give details,
however, it is worth noticing that with our notation $\sJ^-_v$ denotes
simultaneously the left and right currents and the vector and axial
currents (all of them associated to the LR version of the effective
action). The chiral currents are defined by
\begin{equation}
J^-_{v,R}=(\sJ^-_v)_R \,, \quad
J^-_{v,L}=(\sJ^-_v)_L \,,
\end{equation}
so that (consistently with Convention 1)
\begin{equation}
\delta W^-_{\text{LR}}[v,m]= \frac{1}{2}\langle J^-_{v,R}\delta v_R -J^-_{v,L}\delta v_L
\rangle\,.
\end{equation}
On the other hand, the vector and axial currents are defined by
\begin{equation}
\delta W^-_{\text{LR}}[v,m]= \langle  J^-_V\delta V + J^-_A\delta A \rangle\,.
\end{equation}
Thus
\begin{equation}
J^-_{V,A}= \frac{1}{2}(J^-_{v,R} \mp J^-_{v,L}) =\sJ^-_v\,.
\end{equation}
In the last equality we are using our conventions with the proviso
that $J^-_V$ and $J^-_A$ are pseudo-parity odd and even quantities,
respectively. (Of course the usual vector and axial current are those
associated to the VA version of the effective action, so it still
remains to pick up the contribution from the counterterm
$P_{\text{ct}}[v]$.)

\subsection{The two-dimensional pseudo-parity odd effective action 
from the anomaly}
\label{subsec:V.d}

In this subsection we will point out a general property of the
effective action in the pseudo-parity odd sector, which holds to all
orders and for any gauge group and any space-time dimension greater
than zero, and will show that this property is sufficient to
completely fix $W^-[v,m]$ at leading order in two dimensions from the
chiral anomaly.

The general property is that $W^-[v,m]$ vanishes identically when
there are no gauge fields and one of the scalar fields, say $\mrl$, is
a space-time constant, that is,
\begin{equation}
W^-[v,\mlr,\mrl]=0\,,\quad \text{when}\quad v=0, \quad d\mrl=0\,.
\label{eq:n113}
\end{equation} 
To proof this statement, let us consider the variation of $W^-[v,m]$
within this class of configurations when only $\mlr$ is varied. Use of
\eq{n6a} yields
\begin{eqnarray}
\delta W^-[v,m] &=& -\frac{1}{2}\Tr\left[ \gamma_5
\frac{1}{\mlr -\thru\partial \mrl^{-1} \thru\partial} \delta\mlr 
 \right] \nonumber \\
 &=& -\frac{1}{2}\Tr\left[ \gamma_5
\frac{1}{\mlr - \mrl^{-1}\partial^2} \delta\mlr 
\right] \nonumber \\
&=& 0\,.
\end{eqnarray}
The second equality holds due to $d\mrl=0$. The last equality follows
from $\tr\gamma_5=0$ (except at $d=0$, and indeed the property does
not hold in this case). Therefore the value of
$W^-[v=0,\mlr,\mrl=\text{constant}]$ does not depend on $\mlr$. This
value is zero as follows from choosing $\mlr=\mrl$, since in this case
the configuration is unchanged under pseudo-parity conjugation and the
pseudo-parity odd component vanishes. Note that this property is
specific of the effective action functional and does not derive from
general symmetry properties of this functional.  From \eq{n36}, it
follows that within this class of configurations
\begin{equation}
W^-_c[v,m] =  -\Gamma_{\text{gWZW}}[v,m]\,,\quad (v=0, \quad d\mrl=0)\,.
\label{eq:n36a}
\end{equation}
All higher orders in the derivative expansion must vanish separately,
whereas the leading term of $W^-_c[v,m]$ must cancel the extended
gauged WZW term.

Next we will use this property to determine the chiral covariant
remainder in two dimensions. To do this let us compute the two sides
of \eq{n36a} when $v=0$, $\mrl=1$ (or any constant c-number) and
$\mlr=\mu$ (this is just a change of name).  Using
$(\sm^{-1}d\sm)_R=\mu^{-1}d\mu$, and $(\sm^{-1}d\sm)_L=0$, one finds
\begin{equation}
\Gamma_{\text{gWZW}}[v,m]=\frac{1}{2}\Gamma_{\text{LR}}[v=0, U=\mu]
=\frac{1}{2}\left\langle -\frac{1}{3} (\mu^{-1}d\mu)^3\right\rangle
=\frac{1}{2}\left\langle h_{\text{WZW}}(\mu_1,\mu_2)d\mu^2\right\rangle\,,
\end{equation}
where the function $h_{\text{WZW}}(z_1,z_2)$ was introduced in
\eq{n25f}.

On the other hand, using only symmetry arguments (including
analyticity of the effective action functional) the leading term of
$W^-_c[v,m]$ in two dimensions must have the form given in \eq{n14}
with
\begin{equation}
N(\sm_1,\sm_2)= \sm_1\sm_2\,n(\sm_1^2,\sm_2^2) \,,
\end{equation}
for some antisymmetric function $n(z_1,z_2)$ to be determined. For the
class of configurations selected above, and using that in this case
$m_L^2=m_R^2=\mu$, $(\sm d\sm)_R=d\mu$, $(\sm d\sm)_L=0$, one finds
\begin{equation}
W^-_c[v,m] =  \frac{1}{2}\left\langle
n(\mu_1,\mu_2)d\mu^2\right\rangle \,.
\end{equation}
Comparing both calculations, it follows that $n(z_1,z_2)=
-h_{\text{WZW}}(z_1,z_2)$ and thus
\begin{equation}
N(\sm_1,\sm_2)= -\sm_1\sm_2 h_{\text{WZW}}(\sm_1^ 2,\sm_2^ 2)\,,
\label{eq:n119}
\end{equation}
which is indeed verified by the correct function $N_{12}$ given in
\eq{n13a}.

The points to remark are i) since the WZW term is completely
determined by integration of the chiral anomaly, the function
$h_{\text{WZW}}(z_1,z_2)$ also follows from the anomaly, ii) although
$W^-_c$ is considered for a particular case, this is sufficient to
determine the function $N_{12}$ because no special properties of
$\mu$ (i.e. particular flavor groups) have been assumed.

In four dimensions this method is insufficient to fix the effective
action. The function $N_{123}$ does not contribute since $\sF=0$ when
we take $v=0$. On the other hand, $N_{1234}$ can be decomposed as
\begin{eqnarray}
N_{1234} &=& 
n_{1234}
+\sm_1\sm_2\,n'_{1234}
+\sm_2\sm_3\,n'_{2341}
+\sm_3\sm_4\,n'_{3412}
-\sm_1\sm_4\,n'_{4123}
\nonumber \\ && 
+\sm_1\sm_3\,n''_{1234}
+\sm_2\sm_4\,n''_{2341}
+\sm_1\sm_2\sm_3\sm_4\,n'''_{1234} \,,
\end{eqnarray}
where the various $n$'s are functions of $\sm^2$. For configurations
with $v=0$ and $\mrl=1$, only the last component $n'''_{1234}$ gives a
contribution, hence all other components remain undetermined by this
procedure. The component $n'''_{1234}$ is fixed by imposing
cancellation with the contribution coming from
$\Gamma_{\text{gWZW}}[v,m]$. We have explicitly verified this with our
formulas. In passing, we note another unexpected property, namely, the
component $n_{1234}$ vanishes identically, although this is not
required by the general symmetries of $N_{1234}$. It seems to be a
specific property of the effective action functional (this statement
depends only of our choice of $\Gamma_{\text{gWZW}}[v,m]$). The other
components do not vanish identically.

To finish this subsection, let us note that the same observation and
method described above can be adapted to the VA version of the
effective action. The fact that $W^-_{\text{LR}}[v,m]$ vanishes when
$v=0$ and $\mrl$ is a constant, \eq{n113}, implies (making a chiral
rotation to the case $P=0$ and using the formulas in Appendix
\ref{app:A}) that
\begin{equation}
W^-_{\text{VA}}[v,m]=-\Gamma_{\text{WZW}}(U)\,,\quad
\text{when}\quad v_{R,L}= U^{\pm 1/2}dU^{\mp 1/2}\,,\quad
\mlr=\mrl=U\,.
\end{equation}
This identity can then be used to determine the function $M_{12}$ in
$W^-_{\text{VA}}[v,m]$ in two dimensions. A straightforward
calculation yields
\begin{equation}
M(z_1,z_2)= \frac{4 z_1 z_2}{z_1+z_2}h_{\text{WZW}}(z_1,z_2)\,.
\end{equation}
This relation is verified by the correct function $M_{12}=A_{12}$.
Combining this formula and that in \eq{n119} yields
\begin{equation}
N(\sm_1,\sm_2)= 
-\frac{\sm_1^2+\sm_2^2}{4\sm_1\sm_2}A(\sm_1^ 2,\sm_2^ 2)\,,
\end{equation}
which is a non-trivial relation between the covariant current,
$A_{12}$, and the covariant remainder, $N_{12}$.

\subsection{Further properties of the extended gauged WZW action}
\label{subsec:V.c}

The functional $\Gamma_{\text{gWZW}}[v,m]$ in \eq{n36} is required to
reproduce the correct chiral anomaly but otherwise it is a matter of
choice. A different choice would be compensated by a change in the
chiral invariant remainder. Nevertheless, the concrete form proposed
in \eq{n37} is the unique such functional enjoying two further
properties, namely, i) it does not mix $\mlr$ with $\mrl$, and ii) it
is invariant under the transformation $\sm\to\sm^{-1}$ (i.e.
$\mlr\leftrightarrow\mrl^{-1}$). The second property is manifest in
\eq{n15} for the gauged terms. For the WZW term it holds too:
\begin{equation}
\left\langle\sR^{d+1}\right\rangle \to
\left\langle\sL^{d+1}\right\rangle =
\left\langle(-\sm\sR\sm^{-1})^{d+1}\right\rangle = -\left\langle
\sm\sR^{d+1}\sm^{-1}\right\rangle = \left\langle
\sR^{d+1}\right\rangle \,.
\end{equation}

That $\Gamma_{\text{gWZW}}[v,m]$ is fully characterized by these two
properties can be seen after a detailed analysis: any other such
functional would differ by a chiral invariant contribution, of the
same form as $W^-_c[v,m]$ in \eq{n14}. The requirement of not mixing
$\mlr$ and $\mrl$ only allows $a\langle\sR_c^4\rangle$ for the term
with $N_{1234}$ (e.g., a piece $\sm^2$ introduces a mixing, and
similarly $\sm\sm'$ or $\sm'^2$), and such a term vanishes
identically. For the term with $N_{123}$, the most general form not
mixing $\mlr$ and $\mrl$ would be $\langle
a\sR_c^2\sF+b\sL_c^2\sF\rangle$, however, mirror symmetry requires
$b=-a$ and this in conflict with invariance under $\sm\to\sm^{-1}$,
which requires $b=a$.

As noted, the property of not mixing $\mlr$ and $\mrl$ does not extend
to the full effective action. Let us discuss the property of
invariance under $\sm\to\sm^{-1}$. First of all, note that it cannot
be a symmetry of the effective action beyond the leading term in the
derivative expansion, since it does not preserve the dimensional
counting, so our next comments refer to this leading term only (for
$W^-$ or the term with precisely $d$ derivatives for $W^+$).

On the chiral circle, the transformation $\sm\to\sm^{-1}$ is a trivial
symmetry (since $\sm^2=1$). As we have just seen, it is also a
symmetry of the functional $\Gamma_{\text{gWZW}}[v,m]$ (on or off the
chiral circle).  Remarkably, it turns out to be an invariance of the
leading term of $W^-[v,m]$ in zero and two dimensions. In the zero
dimensional case this is obvious since $W^-_c[v,m]$ vanishes. In two
dimensions it is an accidental symmetry which follows as an automatic
consequence of chiral and Lorentz invariance, plus scale invariance
and mirror symmetry. Indeed, the most general form of $N_{12}$
consistent with scale invariance and mirror symmetry is
\begin{equation}
N(\sm_1,\sm_2)= \frac{1}{\sm_1\sm_2}f(\sm_1\sm_2^{-1}),\quad
f(x)= - f(x^{-1})\,.
\end{equation}
On the other hand, under the transformation $\sm\to\sm^{-1}$,
$\sm'\to-\sm^{-1}\sm'\sm^{-1}$ and so 
\begin{equation}
N(\sm_1,\sm_2)\to
-\frac{1}{\sm_1^2\sm_2^2}N(\sm_1^{-1},\sm_2^{-1}) = N(\sm_1,\sm_2)\,.
\end{equation}
The same invariance is also automatic in the term with two covariant
derivatives in $W^+$ in two dimensions (although it fails in zero
dimensions for $W^+$). In four dimensions such as a transformation is
not a symmetry of the leading term of $W^-[v,m]$, as can be seen using
the explicit formula of $N_{123}$ in Appendix \ref{app:B} or
$A_{123}$ for the effective current.

\subsection{General form of the chiral invariant remainder}
\label{subsec:V.b}

In Section \ref{subsec:IV.b} we have noted that the forms taken in
\eq{n14} for $W^-_c[v,m]$ in two and four dimensions are actually the
most general ones form those functionals. To show this, let us begin
by considering a functional of the form $\langle N_1\sF\rangle$ in two
dimensions. Using the identity
\begin{equation}
0= \langle(f_1\sm')'\rangle=
\langle(\Delta f)_{12}\sm'^2+f_1\sm''\rangle=
\langle(\Delta f)_{12}\sm'^2+f_1[\sF,\sm]\rangle=
\langle(\Delta f)_{12}\sm'^2 -2\sm f_1\sF\rangle\,,
\end{equation}
it follows that $\langle N_1\sF\rangle$ can be reabsorbed in $\langle
N_{12}\sm'^2\rangle$ by taking $f_1=-\frac{1}{2}N_1$.

In four dimensions, using $\sm''=[\sF,\sm]$ and $\sF'=0$, the most
general form is that given in \eq{n14} augmented with terms of the
form $\langle N'_{12}\sF^2\rangle$. However, due to mirror symmetry
\begin{equation}
N'_{12}=-N'_{21}\,,\quad N'_{12}=(\sm_1-\sm_2)n''_{12} \,,
\end{equation}
that is, the function $n''_{12}=N'_{12}/(\sm_1-\sm_2)$ is finite (in
the coincidence limit) if $N'_{12}$ is finite. Then,
\begin{equation}
\langle N'_{12}\sF^2\rangle=
\langle n''_{12}[\sm,\sF]\sF\rangle=
\langle -n''_{12}\sm''\sF\rangle=
\langle (\Delta n)_{123}\sm'^2\sF +(-n''_{12}\sm'\sF)'\rangle
= \langle (\Delta n)_{123}\sm'^2\sF\rangle\,.
\end{equation}
Therefore a term $\langle N'_{12}\sF^2\rangle$ is also redundant.

Let us now discuss the existence of more general chiral invariant
functionals which do not have the analytical form in \eq{n14}. Since
the functional is chiral invariant it can be computed in a chirally
rotated configuration.  The point is that it is always possible to
chirally rotate a configuration so that $\mrl=\mlr=S$, $P=0$. In the
chiral gauge $P=0$ the only remaining freedom is that of vector gauge
transformations. Therefore, there are as many chiral invariant functionals
of $\mlr$, $\mrl$, $v_R$ and $v_L$ as there are vector gauge invariant
functionals of $S$, $V$ and $A$. These rotated VA fields depend on the
original chiral fields in a non-analytical way.

The most general VA functional has been considered in Section
\ref{subsec:V.f}, \eq{n134}. In four dimensions (and assuming mirror
symmetry) it depends on four independent functions, $M_{123}$,
$M_{1234}$, $M'_{1234}$ and $M''_{123}$. When the functional derives
from an analytical form, these functions take the form given in
\eq{145}. In particular $M_{123}$, and $M_{1234}$, coincide with the
functions $A_{123}$ and $A_{1234}$ of the effective currents. (The
gWZW contribution has to be removed from theses functions but this
does not change the argument.) Since the current determines the
effective action, it follows that $M_{123}$ and $M_{1234}$ determine
$N_{123}$ and $N_{1234}$, and so determine the other two functions
$M'_{1234}$ and $M''_{123}$. This already implies that the analytical
form is not the most general one, since one could imagine new
functionals obtained by keeping the same $M_{123}$ and $M_{1234}$ but
arbitrarily modifying $M'_{1234}$ and $M''_{123}$.  Such functionals
would no be equivalent to an analytical one for any choice of
$N_{123}$ and $N_{1234}$.

Even in two dimensions, where the VA functional contains only one
arbitrary function $M_{12}$, the analytical functional $\left\langle
N_{12}\sm'^2 \right\rangle$ is not the most general one. When the VA
functional is analytical (in terms of the unrotated variables)
\begin{equation}
M_{12}=-2(S_1+S_2)N_{12} \,.
\end{equation}
(This is just \eq{n12} removing the gWZW contribution. $N_{12}$ is
evaluated at $\sm_{1,2}=S_{1,2}$.) The function $N_{12}$ is restricted
by consistency, cyclic symmetry and finiteness (we do not enforce
mirror symmetry here) and this implies
\begin{equation}
M_{12}=-M_{\m{1}\m{2}}\,,\quad
(S_1-S_2)M_{12}=(S_1+S_2)M_{\m{2}1} \,,\quad
M_{1\m{1}}=0\,.
\end{equation}
If $M_{12}$ is analytical in $S$, the first condition follows from
dimensional counting (unless the VA functional breaks scale invariance
or depends on new external fields), but the other two conditions are
not required to have an acceptable VA functional ($M_{12}$ still has
to be finite at $S_1=S_2$). For instance
\begin{equation}
\Gamma[S,V,A]= \left\langle\frac{1}{S}S'A\right\rangle
\end{equation}
violates the conditions and so it cannot be written as $\left\langle
N_{12}\sm'^2 \right\rangle$ for some suitable $N_{12}$.

Another comment is the following. At the end of Section
\ref{subsec:II.B} we noted that one could consider phenomenological
contributions of the form $\langle h(u_1,u_2)du^2\rangle$ in two
dimensions (and similar comments apply to four dimensions as well)
which are consistent with vector gauge invariance but are not chiral
invariant except when the true function $h_{\text{WZW}}(u_1,u_2)$ is
used. Chiral invariance in no longer a problem for functionals of the
form $\langle h(\sm_1,\sm_2)\sm'^2\rangle$ (i.e. the same form of
$W^-_c[v,m]$ but with a different function). Such phenomenological
terms, which are vanishing on the chiral circle, are topological in
the sense that they do not contribute the strength-energy tensor and
their corresponding baryonic current is conserved independently of the
equations of motion. This is can be seen from \eq{112}: setting $v=0$
and taking the trace it says that the baryonic current is a closed
form, and this result does not depend of the explicit form of
$W^-_c[v,m]$.

\subsection{Descent relations}
\label{subsec:V.h}

It is known that the VA version of pseudo-parity odd component of the
effective action equals $2\pi i$ times the baryon number in two more
dimensions \cite{D'Hoker:1984ph,Ball:1986qr,Ball:1989xg} (see
\cite{Salcedo:1996qy} for a proof in the framework of the
$\zeta$-function regularized effective action). In this relation one
of the extra dimensions, $u$, is regarded as the time and the other,
$v$, is a new space direction. The relation holds provided that the
dependence of the $d+2$-dimensional configuration is $u$-independent
and adiabatic in $v$, so that no more than one $v$-derivative is
retained. In this case, and choosing $\eta_d=i^{d/2}$, the relation
takes the form
\begin{equation}
W^-_{\text{VA},d}
=-2\pi i\langle \sJ^-_{{\text VA},V} \rangle_{d+2}\,,
 \quad(\eta_d=i^{d/2})\,.
\label{eq:n154}
\end{equation}
The subscripted dimension in the right-hand side refers to the
normalization of $\langle\ \rangle$, \eq{n2}. $\sJ^-_{{\text VA},V}$
denotes the vector current associated to
$W^-_{\text{VA,d+2}}[v,m]$. Under the conditions stated above, and due
to gauge invariance, only the pseudo-parity odd component of the
current has a contribution to the baryon number
\cite{Salcedo:1996qy}.

The previous relation can be rewritten as one for the LR version as
follows
\begin{equation}
W^-_d[v,m] =-2\pi i\langle \sJ^-_{v,c} \rangle_{d+2}
-2W_{\text{CS},d+1}[v] \,,
\label{eq:n155}
\end{equation}
where $W_{\text{CS}}$ is the Chern-Simons action
\begin{eqnarray}
W_{\text{CS},d=1}[v] &=& 
 \left\langle \sv \right\rangle_{d=0} \nonumber  \\ 
W_{\text{CS},d=3}[v] &=& 
 \left\langle  \frac{1}{3}\sv^3 - \sv\sF   \right\rangle_{d=2} \,.
\end{eqnarray}
The Chern-Simons terms are precisely those appearing in \eq{n15}, and
account for all the chiral symmetry breaking in $W^-_d[v,m]$. (The
factor of 2 in $W_{\text{CS}}$ accounts for a Chern-Simons term for
the right field and another for the left field.)

Let us detail the derivation of \eq{n155} for $d=2$ (assuming
\eq{n154}). The four-dimensional counterterm relating the VA and LR
versions is given in \eq{A7}. Its contribution to the vector current
is minus
\begin{equation}
J_{\text{ct},d=4}=
-2\{\sF,\sv\}+2\sv^3 +6d(v_Rv_L)\,.
\end{equation}
This contribution is to be combined in \eq{n154} with that of the
counterterm current, relating the consistent and covariant currents,
\eq{30}. This yields
\begin{equation}
2\pi i\left\langle\sP - J_{\text{ct}}\right\rangle_{d=4} =
2W_{\text{CS},d=3} + P_{\text{ct},d=2} \,
\end{equation}
from which \eq{n155} follows.

A we have said, the chiral breaking terms coincide at both sides of
\eq{n155}. On the other hand, equating the chiral preserving terms at
both sides gives a relation between the functions $N$ in $d$
dimensions and the functions $A$ in $d+2$
dimensions. ($\sJ^-_{v,c,d+2}$ is a $d+1$-form, so $W^-_d[v,m]$ must
first be brought to a $d+1$-dimensional form by applying $\sDa$.)  For
$d=0$ and $d=2$ the relations are
\begin{equation}
0 = f_1 = f_{12}-f_{2\m{1}} = f_{123}-f_{23\m{1}}+f_{3\m{1}\m{2}} \,,
\end{equation}
where
\begin{eqnarray}
f_1 &=& \frac{1}{\sm_1} +\frac{1}{2}A_{1\m{1}}\,,
\nonumber \\
f_{12} &=& 
\frac{1}{\sm_1} + \frac{1}{\sm_2}
-2(\sm_1 -\sm_2)N_{12} 
- \frac{1}{6}(A_{12\m{1}} +A_{21\m{2}}) \,,
\nonumber \\
f_{123} &=& 
-\frac{1}{3}\frac{1}{\sm_1\sm_2\sm_3}
+(\Delta N)_{123}
- \frac{1}{6}A_{123\m{1}} \,.
\end{eqnarray}
These relations are checked by our calculation.

\section*{Acknowledgments}
This work was supported in part by funds provided by the Spanish
DGICYT grant no. PB98-1367 and Junta de Andaluc\'{\i}a grant
no. FQM-225.

\appendix

\section{Chiral anomaly and WZW action}
\label{app:A}

Here we will collect some formulas which are needed in the text. The
variation of the effective action under infinitesimal chiral rotations
is the (consistent) chiral anomaly. As is well known, $W^+$ can be
renormalized so that it is free from chiral anomalies and hence only
the pseudo-parity odd component of the effective action is necessarily
anomalous.

Let $\Omega_{L,R} =\exp(\alpha_{R,L})$. Then, the LR version of the
anomaly takes the form (where $\alpha$ is infinitesimal)
\begin{eqnarray}
\delta W^-_{\text{LR},d=0}[v,m] &=& -\left\langle \alpha_R-\alpha_L
\right\rangle
=  \left\langle -2\alpha\right\rangle \,, \nonumber \\
\delta W^-_{\text{LR},d=2}[v,m] &=& \left\langle
v_R d\alpha_R-v_L d\alpha_L\right\rangle
= \left\langle 2 \sv d\alpha \right\rangle \,,
\label{eq:27} \\
\delta W^-_{\text{LR},d=4}[v,m] &=&
\left\langle \left( -4\sF\sv+2\sv^3 \right)d\alpha \right\rangle
\,.  \nonumber
\end{eqnarray}
The LR anomaly presents two key features, first it does not depend on
$m$ and second, the two chiral sectors do not mix. In addition, it is
consistent, i.e. a true variation. Let $(v,m)$ be a field
configuration obtained from another configuration $(\bar{v},\bar{m})$
through a chiral rotation $(\Omega_L,\Omega_R)$,
i.e. $(v,m)=(\bar{v},\bar{m})^\Omega$. Then integration of the anomaly
yields
\begin{equation}
W^-_{\text{LR}}[v,m]= W^-_{\text{LR}}[\bar{v},\bar{m}] +
\Gamma[v_R,\Omega_R]-\Gamma[v_L,\Omega_L] \,.
\label{eq:28}
\end{equation}
($\Gamma[v,\Omega]$ is the same function in both cases but with different
arguments.) Reflecting the same property of the LR anomaly, the
variation is composed of two terms which are not mixed and are
independent of $m$. Explicitly,
\begin{eqnarray}
\Gamma_{d=2}[v,\Omega] &=& 
\left\langle
-\frac{1}{3}(r_c^3 +v^3) + ( r_c  + v ) F 
\right\rangle  \nonumber \\
&=& \left\langle
-\frac{1}{3}r^3 +vr \right\rangle \,, \\
\Gamma_{d=4}[v,\Omega] 
&=& \left\langle
-\frac{1}{5}(r_c^5 +v^5)
+(r_c^3+ v^3)F
-2 (r_c + v)F^2
 \right\rangle \,,
\nonumber \\
&=& \left\langle
-\frac{1}{5}r^5
+vr^3 +v^2r^2-\frac{1}{2}(vr)^2 + v^3r -2Fvr
 \right\rangle \,,
\end{eqnarray}
where $r= \Omega^{-1}d\Omega$ and $r_c= \Omega^{-1}d\Omega - v$.

The VA version of the effective action is characterized by being
vector gauge invariant. It is obtained from the LR version by
subtracting an appropriate local polynomial counterterm,
\begin{equation}
W^-_{\text{VA}}[v,m] = W^-_{\text{LR}}[v,m] - P_{\text{ct}}[v]\,.
\label{eq:29}
\end{equation}
Note that the counterterm is independent of $m$. Explicitly
(Convention 1 applies)
\begin{eqnarray}
P_{\text{ct},d=2}[v] &=&  \left\langle v_Rv_L \right\rangle \,,
\nonumber \\
P_{\text{ct},d=4}[v] &=& 
\left\langle 
2F_R[v_L,v_R]
+2v_Rv_L^3 
-\frac{1}{2}(v_Rv_L)^2
\right\rangle \,.
\label{eq:A7}
\end{eqnarray}
The corresponding VA anomaly is thus
\begin{eqnarray}
\delta W^-_{\text{VA},d=2}[v,m] &=&
\left\langle 
4\left(F_V-A^2\right)\alpha_A
\right\rangle \,, \\
\delta W^-_{\text{VA},d=4}[v,m] &=&
\left\langle 
-4\left(3F_V^2+F_A^2-4AF_VA-\{F_V,A^2\}-A^4\right)\alpha_A
\right\rangle \,, \nonumber
\end{eqnarray}
where $v_{R,L}=V\pm A$, $F_V=D_V^2=dV+V^2$ and $F_A=\{D,A\}$. In
addition, we have introduced the vector and axial variations through
$\alpha_{R,L}=\alpha_V\pm\alpha_A$. As advertised, in this case there
is no anomaly associated to vector transformations.

Consider now the variation of the VA effective action
\begin{equation}
W^-_{\text{VA}}[v,m]= W^-_{\text{VA}}[\bar{v},\bar{m}]
+\Gamma_{\text{VA}}[v,U]\,,\quad U:=\Omega_L^{-1}\Omega_R\,.
\end{equation}
$\Gamma_{\text{VA}}[v,U]$ is the gauged WZW action which, by
construction, saturates the VA anomaly. Because the anomaly is
independent of $m$, so is $\Gamma_{\text{VA}}[v,U]$. In addition,
since $W^-_{\text{VA}}$ is vector gauge invariant its variation
depends on $\Omega_{L,R}$ only through the combination
$U=\Omega_L^{-1}\Omega_R$, i.e., the axial part of $\Omega$. The LR
form of this relation is obtained by adding the counterterm
$P_{\text{ct}}[v]$. This gives
\begin{eqnarray}
W^-_{\text{LR}}[v,m] &=& 
W^-_{\text{VA}}[\bar{v},\bar{m}] +\Gamma_{\text{LR}}[v,U] \,.
\label{eq:26}
\end{eqnarray}
Note that by construction $\Gamma_{\text{VA}}[v,1]=0$ and
$\Gamma_{\text{LR}}[v,1]= P_{\text{ct}}[v]$, so
\begin{equation}
\Gamma_{\text{VA}}[v,U]= \Gamma_{\text{LR}}[v,U]-\Gamma_{\text{LR}}[v,1]\,.
\end{equation}
$\Gamma_{\text{LR}}[v,1]$ is known as the Bardeen subtraction.

Comparing eqs.~(\ref{eq:28}), (\ref{eq:29}) and (\ref{eq:26}), it
follows that
\begin{equation}
\Gamma_{\text{LR}}[v,U] = 
\Gamma(v_R,\Omega_R)-\Gamma(v_L,\Omega_L) +
P_{\text{ct}}[\bar{v}] \,.
\end{equation}
On the other hand, noting that the Bardeen subtraction vanishes for
purely right or left gauge fields, yields
\begin{equation}
\Gamma[v,\Omega] = \Gamma_{\text{LR}}[v_R=v,v_L=0,U=\Omega]\,.
\end{equation}
Explicitly, in two dimensions
\begin{eqnarray}
\Gamma_{\text{LR},d=2}[v,U] =
\left\langle
-\frac{1}{3}(U^{-1}dU)^3
-U^{-1}dU v_R + UdU^{-1} v_L
-U^{-1}v_LUv_R \right\rangle\,.
\label{eq:A13}
\end{eqnarray}
In order to use the Conventions 1 and 2, let us define $\sU$ as
$(\sU)_{LR}=U$ and $(\sU)_{RL}=U^{-1}$. Note that $\sU^{-1}$ equals
$\sU$ with our conventions. In addition, let $\sR=\sU d\sU$. Then
\begin{eqnarray}
\Gamma_{\text{LR},d=2}[v,U] =
\left\langle
-\frac{1}{3}\sR^3
-2\sR\sv - \sU\sv\sU\sv 
\right\rangle\,.
\end{eqnarray}
In four dimensions
\begin{eqnarray}
\Gamma_{\text{LR},d=4}[v,U] &=& 
\left\langle
-\frac{1}{5}\sR^5 
-2\sR^3\sv
+(\sR\sv)^2
+2\sR^2\sv\sU\sv\sU
+2\sR\sU\sv\sU d\sv
\right. \nonumber \\ && \left.
+2\sR\sv^3
+2\sR\sv\sU\sv\sU\sv
+2(\sR + \sU\sv\sU ) \{\sv,d\sv\}
+2\sU\sv\sU\sv^3
+\frac{1}{2}(\sU\sv\sU\sv)^2
 \right\rangle \,.  \nonumber 
\end{eqnarray}

\section{Explicit formulas for the functions $A$ and $N$ in two and 
four dimensions}
\label{app:B}

For the currents we give the formulas for the associated functions
$\bar{A}$ which are more symmetric. In two dimensions
\begin{eqnarray}
\bar{A}_{12} &=& -\frac{2}{\sm_1+\sm_2} - \frac{2\sm_1\sm_2}
{(\sm_1+\sm_2)(\sm_1^2-\sm_2^2)}\log(\sm_1^2/\sm_2^2) \,.
\end{eqnarray}
In four dimensions
\begin{eqnarray}
\bar{A}_{123}&=& \bar{A}^R_{123} + \bar{A}^L_{123} \log(\sm_1^2/\sm_3^2) 
 + \bar{A}^L_{213} \log(\sm_2^2/\sm_3^2) \, 
\label{eq:69b}\\
\bar{A}_{1234}&=& \bar{A}^R_{1234}
 + \bar{A}^L_{1234} \log(\sm_1^2) 
 + \bar{A}^L_{2341} \log(\sm_2^2) 
 + \bar{A}^L_{3412} \log(\sm_3^2) 
 + \bar{A}^L_{4123} \log(\sm_4^2)
\end{eqnarray}
(where the superindices $R$ and $L$ refer to rational and logarithmic
components, respectively). With
\begin{eqnarray}
\bar{A}^R_{123} &=&
\frac{6(\sm_1\sm_2+\sm_1\sm_3+\sm_2\sm_3)}
{(\sm_1+\sm_2)(\sm_1+\sm_3)(\sm_2+\sm_3)} 
\\
\bar{A}^L_{123} &=&
\frac{6\sm_1^3(\sm_1\sm_2+\sm_1\sm_3+2\sm_2\sm_3)}
 {(\sm_1+\sm_2)(\sm_1+\sm_3)(\sm_1^2-\sm_2^2)(\sm_1^2-\sm_3^2)}
\,, \\
\bar{A}^R_{1234} &=&
\frac{6(\sm_1\sm_2\sm_3+\sm_1\sm_2\sm_4+\sm_1\sm_3\sm_4+\sm_2\sm_3\sm_4)}
{(\sm_1+\sm_2)(\sm_1+\sm_3)(\sm_1+\sm_4)
(\sm_2+\sm_3)(\sm_2+\sm_4)(\sm_3+\sm_4)} 
\\
\bar{A}^L_{1234} &=&
-\frac{6\sm_1^3
(\sm_1(\sm_2\sm_3+\sm_2\sm_4+\sm_3\sm_4)+2\sm_2\sm_3\sm_4-\sm_1^3)}
 {(\sm_1+\sm_2)(\sm_1+\sm_3)(\sm_1+\sm_4)(\sm_1^2-\sm_2^2)(\sm_1^2-\sm_3^2)
(\sm_1^2-\sm_4^2)} \,.
\label{eq:69c}
\end{eqnarray}

For the effective action in two dimensions
\begin{eqnarray}
N_{12} &=&
-\frac{\sm_1\sm_2}{\sm_1^2-\sm_2^2}\left(\frac{\log(\sm_1^2/\sm_2^2)}
{\sm_1^2-\sm_2^2}-\frac{1}{2}\left(
\frac{1}{\sm_1^2}+\frac{1}{\sm_2^2}\right)\right)
\,.
\end{eqnarray}

In four dimensions, the function $N_{123}$ can be written as
\begin{equation}
N_{123}= N^R_{123}
+N^L_{123} \log(\sm_1^2/\sm_2^2) - N^L_{321} \log(\sm_3^2/\sm_2^2) \,,
\end{equation}
with
\begin{eqnarray}
N^R_{123} &=& 
\frac{1}
{2\sm_1\sm_2\sm_3(\sm_1^2 - \sm_2^2)(\sm_3^2 - \sm_2^2)(\sm_1 - \sm_3)}
\nonumber \\ && \times
\Big(    3\sm_1^2\sm_3^2(\sm_1 - \sm_3)^2 
 +  4\sm_1\sm_2\sm_3(\sm_1 + \sm_3)(2\sm_1^2 - 3\sm_1\sm_3 + 2\sm_3^2-\sm_2^2) 
\nonumber \\ && {\ \ }
 +  \sm_2^2(\sm_1^4 + 10\sm_1^3\sm_3 - 18\sm_1^2\sm_3^2 + 
     10\sm_1\sm_3^3 + \sm_3^4)
 -  \sm_2^4(\sm_1 + \sm_3)^2      \Big)\,,
 \\ 
N^L_{123} &=& 
\frac{2}
{ (\sm_1^2 - \sm_2^2)^2(\sm_1^2 - \sm_3^2)(\sm_1 - \sm_3) }
\nonumber \\ && \times
  \Big(
     \sm_1^4(\sm_2 - 2\sm_3) + \sm_1^2(\sm_2^3 + \sm_3^3) +
  \sm_2^2\sm_3^2(\sm_2 + \sm_3) 
\nonumber \\ &&  {\ \ }
+   \sm_1^3(\sm_2^2 - 3\sm_2\sm_3 - \sm_3^2)  
-  \sm_1\sm_2\sm_3(\sm_2^2 - \sm_3^2)
  \Big) \,.
\end{eqnarray}

Likewise,
\begin{equation}
N_{1234}= N^R_{1234}
+N^L_{1234} \log(\sm_1^2)
+N^L_{234\m{1}} \log(\sm_2^2)
+N^L_{34\m{1}\m{2}} \log(\sm_3^2)
+N^L_{4\m{1}\m{2}\m{3}} \log(\sm_4^2)\,,
\end{equation}
where

\begin{eqnarray}
N^R_{1234} &=& \frac{1}{4}\Big(
	 \frac{2(2\sm_2 + \sm_3)}{
        (\sm_1^2 - \sm_2^2)(\sm_2^2 - \sm_3^2)(\sm_2 - \sm_4)} 
 -    \frac{ 2(2\sm_2 + \sm_1 )}{
(\sm_1^2 - \sm_2^2)(\sm_2^2 - \sm_3^2)(\sm_2 +  \sm_4)}
\nonumber \\ &&
-   \frac{3( \sm_2\sm_3 - \sm_1(\sm_2 + \sm_3)) }{
        \sm_3(\sm_1^2 - \sm_3^2)(\sm_2^2 - \sm_3^2)(\sm_3 - \sm_4)} 
+  \frac{3(\sm_1\sm_2 - \sm_3(\sm_1 + \sm_2))}{
        \sm_1(\sm_1^2 - \sm_2^2)(\sm_1^2 - \sm_3^2)(\sm_1 + \sm_4)}
\nonumber \\ && 
- \frac{\sm_2\sm_3 + \sm_1(\sm_2 + \sm_3)}{
        \sm_1(\sm_1^2 - \sm_2^2)(\sm_1^2 - \sm_3^2)(\sm_1 - \sm_4)} 
 +  \frac{\sm_2\sm_3 + \sm_1(\sm_2 + \sm_3)}{
        \sm_3(\sm_1^2 - \sm_3^2)(\sm_2^2 - \sm_3^2)(\sm_3 + \sm_4)}
\nonumber \\ && 
+   \frac{1}{\sm_1\sm_2\sm_3\sm_4} 
\Big)\,,
\end{eqnarray}
\begin{eqnarray}
N^L_{1234} &=& 
\frac{1}{
     2(\sm_1^2 - \sm_2^2)^2(\sm_1^2 - \sm_3^2)^2(\sm_1^2 - \sm_4^2)^2}
\nonumber \\ && \times
\Big(
     6\sm_1^7\sm_3 + (\sm_2 - \sm_4) (\sm_2^2\sm_3^3\sm_4^2+3\sm_1^6\sm_3)
- \sm_1\sm_2\sm_3^3\sm_4(\sm_2 - \sm_4)^2 + 
\nonumber \\ && 
 +  \sm_1^2\sm_3^2(\sm_2^3(2\sm_4 + \sm_3 ) - \sm_4^3(2\sm_2 + \sm_3))
\nonumber \\ && 
-    \sm_1^4(\sm_2 - \sm_4)(2\sm_2^2(\sm_3 + \sm_4) +
     \sm_2\sm_4(\sm_3 + 2\sm_4) 
+  2\sm_3(\sm_3^2 + \sm_4^2)) 
\nonumber \\ && 
+ \sm_1^3
        (-\sm_3^2\sm_4^3 + \sm_2^2\sm_4^2(2\sm_3 + \sm_4) + \sm_2\sm_3\sm_4(2\sm_3^2 + \sm_4^2) + 
          \sm_2^3(-\sm_3^2 + \sm_3\sm_4 + \sm_4^2)) 
\nonumber \\ && 
    -   \sm_1^5(\sm_2^2(4\sm_3 + \sm_4) 
+ \sm_2(-\sm_3^2 + 2\sm_3\sm_4 + \sm_4^2)
+           \sm_3(2\sm_3^2 - \sm_3\sm_4 + 4\sm_4^2))\Big) \,.
\end{eqnarray}

We have tried to write these formulas in a form as simple as
possible. In the case of $N^R_{1234}$ this implies that the cyclic
property is not manifest.

\end{document}